\documentclass[12pt,letterpaper,onecolumn]{article}

\usepackage[cmex10]{amsmath}
\usepackage{graphicx}
\usepackage{amssymb}
\usepackage{float}
\usepackage{array}
\usepackage{todonotes}
\usepackage{setspace}
\usepackage{upgreek}
\usepackage{cite}
\usepackage{xr}
\usepackage{bm}
\usepackage{multirow}
\usepackage{subfig}
\usepackage{colortbl}

\makeatletter
  \renewcommand\@biblabel[1]{#1.}     
  \def\@cite#1#2{({#1\if@tempswa , #2\fi})}
  \def\eqref#1{[\ref{#1}]}
\makeatother
\makeatletter
  \def\tagform@#1{\maketag@@@{[#1]\@@italiccorr}}
\makeatother

\DeclareMathOperator*{\argmin}{arg\,min}

\usepackage{todonotes}
\begin{document}

\begin{center}
{\huge \bf Robust Autocalibrated Structured Low-Rank EPI Ghost Correction}
\end{center}
 
\vfill

\noindent Rodrigo A. Lobos$^{1,2}$, W. Scott Hoge$^{4,5}$, Ahsan Javed$^{1,2}$, Congyu Liao$^{5,6}$, Kawin Setsompop$^{5,6}$, Krishna S. Nayak$^{1,2,3}$, Justin P. Haldar$^{1,2,3}$

\vfill

\noindent $^1$Ming Hsieh Department of Electrical and Computer Engineering, University of Southern California, Los Angeles, CA, USA

\noindent $^2$ Signal and Image Processing Institute, University of Southern California, Los Angeles, CA, USA

\noindent $^3$Department of Biomedical Engineering, University of Southern California, Los Angeles, CA, USA

\noindent $^4$Department of Radiology, Brigham and Women's Hospital, Boston, MA, USA

\noindent $^5$Department of Radiology, Harvard Medical School, Boston, MA, USA

\noindent $^6$ Athinoula A. Martinos Center for Biomedical Imaging, Charlestown, MA, USA

\vfill 

\begin{singlespace}
\noindent {\bf Address Correspondence to:}
\vspace{.25in}

\noindent Rodrigo A. Lobos

\noindent University of Southern California

\noindent University Park Campus, 3710 McClintock Avenue, Ronald Tutor Hall (RTH) \#317,

\noindent Los Angeles,  CA 90036.

\noindent Tel: \noindent (+1)323-561-2265

\noindent Email: \noindent rlobos@usc.edu
\end{singlespace}

\vfill

\noindent Submitted to Magnetic Resonance in Medicine

\noindent MANUSCRIPT BODY APPROXIMATE WORD COUNT = 6000

\newpage

\section*{ABSTRACT}

\noindent \textbf{Purpose:} 
We propose and evaluate a new structured low-rank method for EPI ghost correction called Robust Autocalibrated LORAKS (RAC-LORAKS).  The method can be used to suppress EPI ghosts arising from the differences between different readout gradient polarities and/or the differences between different shots.  It does not require conventional EPI navigator signals, and is robust to imperfect autocalibration data.

\noindent \textbf{Methods:} 
Autocalibrated LORAKS is a previous structured low-rank method for EPI ghost correction that uses GRAPPA-type autocalibration data to enable high-quality ghost correction.  This method works well when the autocalibration data is pristine, but performance degrades substantially when the autocalibration information is imperfect.  RAC-LORAKS generalizes Autocalibrated LORAKS in two ways.  First, it does not completely trust the information from autocalibration data, and instead considers the autocalibration and EPI data simultaneously when estimating low-rank matrix structure.  And second, it uses complementary information from the autocalibration data to improve EPI reconstruction in a multi-contrast joint reconstruction framework.  RAC-LORAKS is evaluated using simulations and in vivo data, including comparisons to state-of-the-art methods.

\noindent \textbf{Results:} 
RAC-LORAKS is demonstrated to have good ghost elimination performance compared to state-of-the-art methods in several complicated EPI acquisition scenarios (including gradient-echo brain imaging, diffusion-encoded brain imaging, and cardiac imaging).

\noindent \textbf{Conclusion:} 
RAC-LORAKS provides effective suppression of EPI ghosts and is robust to imperfect autocalibration data.  

\section*{KEYWORDS}
Echo-planar imaging; Nyquist ghost correction; structured low-rank matrix recovery; constrained reconstruction; multi-contrast reconstruction.

\newpage 

\section{INTRODUCTION}
Echo-planar imaging (EPI) is a widely-used high-speed MRI acquisition strategy \cite{stehling1991},  but is subject to several undesirable artifacts \cite{bernstein2004}.  Nyquist ghosts are one of the most common EPI artifacts, and occur because of systematic differences between  the interleaved lines of k-space that are acquired with different readout gradient polarities, and/or because of systematic differences between interleaved lines of k-space data that are acquired with different shots in a multi-shot acquisition.   Despite substantial efforts over several decades to solve this problem  \cite{bernstein2004, Bruder_1992,Hu_1996,Buonocore_1997,Chen_2004,skare2006fast,xiang2007correction,hoge2010robust,Xu_2010,hoge2015dual,chang2016,Yarach_2017,Ianni_2017,Xie_2017,Lee_2016,Mani_2016,lobos2018navigator, Lyu_2018, Liu_2019}, the widely-deployed modern ghost correction schemes are still prone to incomplete ghost suppression, as illustrated in Supporting Information Fig.~S1.

Recently, structured low-rank matrix methods for ghost correction  \cite{Lee_2016,Mani_2016,lobos2018navigator, Lyu_2018, Liu_2019} have received increasing attention for their ability to provide excellent ghost-suppression performance without the need for additional  ``navigator'' information (i.e., reference scans collected alongside each EPI readout that allow estimation of the systematic inconsistencies between different gradient polarities or different shots).  These methods can suppress ghosts better than navigator-based methods, and eliminate the need to acquire navigators for each EPI readout. Among  different structured low-rank matrix approaches, a ghost correction method based on Autocalibrated LORAKS (AC-LORAKS) \cite{haldar2015autocalibrated} was previously demonstrated to yield high-quality results across a range of different scenarios \cite{lobos2018navigator}.  To eliminate a fundamental ambiguity in  structured low-rank matrix recovery from uniformly undersampled EPI data \cite{lobos2018navigator}, AC-LORAKS makes use of parallel imaging subspace information estimated from  autocalibration (ACS) data acquired in a pre-scan. This ACS-based approach is similar to standard autocalibrated parallel imaging methods like GRAPPA \cite{griswold2002}, SPIRiT \cite{lustig2010}, and PRUNO \cite{zhang2011}.

While the AC-LORAKS approach to ghost correction generally works well when the ACS data is pristine and well-matched to the EPI data to be reconstructed, there are many situations where experimental conditions (e.g., subject motion, eddy currents, etc.) can lead to artifacts within the ACS data or mismatches between the ACS  and EPI data.  The performance of the  AC-LORAKS ghost correction procedure degrades in the presence of these ACS artifacts and mismatches.   Note that this kind of issue is not unique to AC-LORAKS or to ghost correction, and imperfect ACS/calibration data is a longstanding  and commonly-reported problem for all calibration-based image reconstruction methods \cite{ying2007,sheltraw2012,hoge2015dual,baron2016,talagala2016,holme2019}. For AC-LORAKS ghost correction, imperfect ACS data can be especially troublesome in contexts where the prescan would  be done once before acquiring a long sequence of multiple EPI images (e.g., in BOLD fMRI or diffusion MRI applications), and then used to reconstruct each image in the sequence.  

In this paper, we propose an extension of AC-LORAKS for EPI ghost correction that is more robust to imperfections in the ACS data.  The new method, called Robust Autocalibrated LORAKS (RAC-LORAKS), has two major differences from the previous AC-LORAKS approach.   First, RAC-LORAKS does not completely trust the subspace information learned from the ACS data, but rather uses a novel structured low-rank matrix formulation that learns subspace information jointly from both the (imperfect) ACS data and the EPI data being reconstructed.  To the best of our knowledge, no previous methods have used this kind of approach to address the longstanding issue of imperfect ACS data. And second, RAC-LORAKS uses the ACS data to provide additional complementary information for the reconstruction of the EPI data within a multi-contrast joint reconstruction framework \cite{bilgic2018improving}.  Preliminary accounts of the first strategy were previously reported in recent conferences \cite{lobos2018robust,lobos2018robust_ismrm}, although we have not previously reported the combination with the second strategy. 

\section{THEORY}
Due to space constraints, our descriptions in this paper will assume that the reader is already familiar with the basic physics of EPI.  Readers interested in a more detailed explanation are referred to classic references \cite{stehling1991,bernstein2004}. For simplicity, our description of EPI ghost correction will generally assume that we are correcting ghosts associated with the differences between data acquired with different readout gradient polarities in a single-shot EPI experiment.  However, since the ghost model for bipolar gradients is nearly identical to the ghost model for multi-shot acquisition, the same approach is easily adapted \textit{mutatis mutandis} to multi-shot acquisition with an arbitrary number of shots \cite{lobos2018navigator}.  
\subsection{Background: Structured Low-Rank EPI ghost correction}
Structured low-rank matrix methods for  EPI ghost correction  \cite{Lee_2016,Mani_2016,lobos2018navigator, Lyu_2018, Liu_2019 } can be viewed as an extension of structured low-rank matrix methods for conventional MR image reconstruction \cite{liang1989,zhang2011,uecker2014,shin2014,haldar2014low,Haldar_2015,ongie2016,jin2015a}, and are based on the same underlying theoretical principles.  In particular, it has been shown that when MRI images have limited support, smooth phase variations, multi-channel correlations, or transform-domain sparsity, then the MRI k-space data will be linearly predictable \cite{haldar2019linear}, which means that  convolutional Hankel- or Toeplitz-structured matrices formed from the k-space data will possess low-rank characteristics.  This observation means that MRI reconstruction can be reformulated as structured low-rank matrix recovery.  Importantly, these structured low-rank matrix recovery methods can even be successful in calibrationless scenarios where ACS data or other prior information about the spatial support, phase, or multi-channel sensitivity profiles is not available \cite{shin2014,haldar2014low, Haldar_2015}.

Structured low-rank EPI ghost correction methods combine these  principles with the earlier observation that EPI data acquired from different gradient polarities or different shots can be treated as coming from different effective ``channels'' in a parallel imaging experiment,  where the systematic differences between different polarities or shots lead to  different phase or magnitude modulations of the underlying EPI image \cite{hoge2015dual,hoge2010robust,chang2016,Xie_2017}.  Since structured low-rank methods for conventional image reconstruction automatically account for the unknown sensitivity maps that modulate the underlying image in a parallel imaging experiment, it is reasonable to apply these same types of methods to handle the unknown polarity- or shot-dependent modulations that manifest in EPI ghost correction. 

For the sake of brevity, we will focus the remainder of our review on the AC-LORAKS method for EPI ghost correction \cite{lobos2018navigator}, since our proposed RAC-LORAKS method is a generalization of AC-LORAKS.  The AC-LORAKS method for EPI ghost correction is based on solving the following regularized optimization problem subject to exact data consistency constraints:
\begin{equation}\label{eq:AC_LORAKS}
\begin{split}
\left\{\hat{\mathbf{k}}^+,\hat{\mathbf{k}}^-\right\} =  \argmin_{\left\{{\mathbf{k}}^+,{\mathbf{k}}^-\right\}} \left\| \mathcal{P}_C(\mathbf{k}^+) \mathbf{N} \right\|_F^2  
 +\left\|  \mathcal{P}_C(\mathbf{k}^-) \mathbf{N}\right\|_F^2  + \lambda J_r([\mathcal{P}_S(\mathbf{k}^+)~ \mathcal{P}_S(\mathbf{k}^-)]).
\end{split}
\end{equation}  
In this expression, $\mathbf{k}^+$ and $\mathbf{k}^{-}$ respectively represent the ideal fully-sampled multi-channel Cartesian k-space data for the positive and negative readout gradient polarities; $\mathcal{P}_C(\cdot)$ is the LORAKS operator that maps the k-space data into a structured matrix that should possess low-rank if the multi-channel image possess limited support and/or interchannel parallel imaging correlations;  $\mathcal{P}_S(\cdot)$ is the LORAKS operator that maps the k-space data into a structured matrix that should possess low-rank if the multi-channel image possess limited support, smooth phase, and/or interchannel parallel imaging correlations; the matrix $\mathbf{N}$ comprises an orthonormal (i.e., $\mathbf{N}^H\mathbf{N} = \mathbf{I}$) basis for the nullspace of the matrix $\begin{bmatrix}\mathcal{P}_C(\mathbf{k}_{\mathrm{acs}^+}) \\ \mathcal{P}_C(\mathbf{k}_{\mathrm{acs}^-}) \end{bmatrix}$, where $\mathbf{k}_{\mathrm{acs}}^+$ and $\mathbf{k}_{\mathrm{acs}}^-$ respectively represent the ACS data for the positive and negative readout gradient polarities; $\lambda$ is a regularization parameter; $J_r(\cdot)$ is a regularization penalty that promotes low-rank characteristics; and $\|\cdot\|_F$ denotes the Frobenius norm.  Due to space constraints, this paper will not provide a detailed recipe for implementing the LORAKS operators $\mathcal{P}_C(\cdot)$ and  $\mathcal{P}_S(\cdot)$, and simply note that our implementations for this paper are identical to those that are described in detail in earlier LORAKS papers \cite{haldar2014low,Haldar_2015}.  There are theoretical benefits to choosing a nonconvex low-rank regularization penalty  \cite{lobos2018navigator}, and the previous AC-LORAKS approach for ghost correction  \cite{lobos2018navigator} used the nonconvex function proposed in the original LORAKS paper \cite{haldar2014low} defined by
\begin{equation}\label{Eq:penalty_fn}
J_r(\mathbf{X}) = \min_{\mathbf{Y}} \|\mathbf{X}-\mathbf{Y}\|_F^2 \text{  s.t.  } \mathrm{rank}(\mathbf{Y}) \leq r,
\end{equation}
where $r$ is a user-selected rank parameter, $\mathbf{X}$ is a matrix representing the point at which we are evaluating the function $J_r(\mathbf{X})$, and $\mathbf{Y}$ is an optimization variable of the same size as $\mathbf{X}$. This penalty encourages matrices that have accurate rank-$r$ approximations.

The first two terms appearing on the right hand side of Eq.~\eqref{eq:AC_LORAKS} respectively impose limited support and parallel imaging constraints on the reconstructions of the positive and negative readout gradient polarities.   The constraints that are used in these terms are implicit in the low-rank characteristics of the structured LORAKS matrices, as captured by the nullspace matrix $\mathbf{N}$.  The nullspace matrix is learned in advance from the ACS data, and as a result, there is an implicit assumption that the support and parallel imaging constraints that were valid for the ACS data are also valid for the EPI data to be reconstructed. Note that if the third term were removed from Eq.~\eqref{eq:AC_LORAKS}, then these first two terms would reduce to performing separate PRUNO \cite{zhang2011} or conventional AC-LORAKS \cite{haldar2015autocalibrated} reconstructions of the data from each polarity.    Acquiring ACS/calibration data is relatively fast and easy, and is already a standard part of most modern parallel imaging protocols, so is not very burdensome on the acquisition.  Using ACS data can also be important in this context, since it has been mathematically proven that structured low-rank matrix methods for ghost correction suffer from fundamental ambiguities unless some form of side information is available \cite{lobos2018navigator}.  While other options exist for removing ambiguity (e.g., using SENSE-like \cite{Pruessmann_1999} image-domain constraints \cite{Mani_2016,lobos2018navigator}), it was previously observed that the AC-LORAKS approach (i.e., using GRAPPA-like \cite{griswold2002} Fourier-domain constraints) offered better performance \cite{lobos2018navigator}.

The third term of Eq.~\eqref{eq:AC_LORAKS} couples the reconstruction of the two polarities together, allowing the reconstruction of one polarity to benefit from information from the other polarity, while also introducing phase constraints to allow the reconstruction to benefit from k-space conjugate symmetry characteristics.  In particular, the third term implicitly and automatically imposes  the following constraints whenever they are compatible with the measured data: limited image support, smooth phase, interchannel parallel imaging correlations, and interpolarity correlations.  Notably, these constraints are all imposed implicitly through the nullspace of a structured matrix, and if a given constraint is not compatible with the measured data, then that constraint will automatically not be imposed by the reconstruction procedure \cite{haldar2019linear}.

The ACS data for AC-LORAKS ghost correction has typically  been acquired using the same process used by Dual Polarity GRAPPA (DPG) \cite{hoge2015dual,polimeni2016reducing,hoge2018dual,liao2019phase}. In particular, assuming a parallel imaging acceleration factor of $R$, DPG employs a $2R$-shot EPI prescan.  The data from different shots and different gradient polarities is then rearranged and interleaved to form one fully-sampled ACS dataset comprised only of data acquired with a positive readout gradient polarity ($\mathbf{k}_{\mathrm{acs}}^+$) and another fully-sampled ACS dataset comprised only of data acquired with a negative readout gradient polarity ($\mathbf{k}_{\mathrm{acs}}^-$).  Since this ACS acquisition strategy is based on a multi-shot approach, it  therefore may be prone to ghosting artifacts due to shot-to-shot variations.  In addition, since the ACS data is often acquired only once at the beginning of a long multi-image EPI scan (e.g., in BOLD fMRI or diffusion MRI experiments), the ACS data acquired at the beginning of the experiment may gradually become mismatched with data acquired at later time points due to scanner drift, subject motion, etc. As noted previously, the ghost correction performance of AC-LORAKS can be substantially degraded when there are mismatches between the ACS data and the EPI data to be reconstructed.    Although a pre-processing procedure has been previously developed to correct for shot-to-shot variations in the ACS data for DPG \cite{hoge2015dual}, this approach is not sufficient for the present context.  In particular, this approach undesirably modifies the magnitude and phase characteristics of the ACS data in ways that are  not well-suited for AC-LORAKS, and only addresses ACS artifacts without accounting for mismatches that may exist between the ACS data and the EPI data. 

\subsection{RAC-LORAKS}
Our proposed RAC-LORAKS method is based on solving the following optimization problem
\begin{equation}\label{eq:mRAC_LORAKS}
\begin{split}
\left\{\hat{\mathbf{k}}^+,\hat{\mathbf{k}}^-, \hat{\mathbf{N}}\right\} & =  \argmin_{\left\{\mathbf{k}^+,\mathbf{k}^-, \mathbf{N}\right\}} \left\| \mathcal{P}_C(\mathbf{k}^+) \mathbf{N} \right\|_F^2  
 +\left\|  \mathcal{P}_C(\mathbf{k}^-) \mathbf{N}\right\|_F^2   \\
 & \,\,\,\,\,\,\,\,\,\,\,\,\,\,\,\,\,\,\,\,\,\,\,\,\,\,\,\,\,\,\,\,
 +\eta \left\| \mathcal{P}_C(\mathbf{k}_{\text{acs}}^+) \mathbf{N} \right\|_F^2 +\eta \left\|  \mathcal{P}_C(\mathbf{k}_{\text{acs}}^-) \mathbf{N}\right\|_F^2 \\
 & \,\,\,\,\,\,\,\,\,\,\,\,\,\,\,\,\,\,\,\,\,\,\,\,\,\,\,\,\,\,\,\,+ \lambda J_r\left(\begin{bmatrix}
\mathcal{P}_S(\mathbf{k}^+)  &  \mathcal{P}_S(\mathbf{k}^-)  & \mathcal{P}_S(\mathbf{k}_{\text{acs}}^+)   & \mathcal{P}_S(\mathbf{k}_{\text{acs}}^-) 
\end{bmatrix}\right)
\end{split}
\end{equation}  
subject to exact data consistency constraints on $\mathbf{k}^+$ and $\mathbf{k}^-$ and subject to orthonormality constraints on $\mathbf{N}$ such that $\mathbf{N}^H\mathbf{N} = \mathbf{I}$. This optimization problem involves four user-selected parameters: the regularization parameters $\eta$ and $\lambda$, the rank parameter $r$, and the number of columns $p$  of the matrix $\mathbf{N}$ (which determines the dimension of the approximate nullspace).

Equation~\eqref{eq:mRAC_LORAKS} has two main differences from Eq.~\eqref{eq:AC_LORAKS}.  First, instead of choosing a predetermined value of the approximate nullspace matrix $\mathbf{N}$ that depends only on the ACS data, $\mathbf{N}$ is now an optimization variable that depends on both the ACS data and the EPI data to be reconstructed.   This allows the reconstruction to be more robust against possible imperfections in the ACS data.  The extent to which the ACS data is trusted is controlled by the user-selected parameter $\eta$.  In the limit as $\eta \rightarrow \infty$, the approximate nullspace matrix $\mathbf{N}$ will converge to the fixed matrix from Eq.~\eqref{eq:AC_LORAKS}.  

The second difference is that the final term of Eq.~\eqref{eq:mRAC_LORAKS} now includes structured matrices formed from the ACS data, in addition to the previous structured matrices formed from the EPI data to be reconstructed.  By concatenating the ACS data in this way, we are essentially treating the ACS data in the same way that we would treat additional channels in a parallel imaging experiment.  Although the ACS data may not have the same contrast as the EPI data to be reconstructed, it has previously been shown that treating multi-contrast information like additional channels in a parallel imaging experiment often leads to improved reconstruction performance  \cite{bilgic2018improving}.  While this improvement has been justified empirically, some level of theoretical justification for this approach can be obtained by modeling different image contrasts as different modulations of some latent  image  \cite{haldar2019linear}.

Algorithmically, Eq.~\eqref{eq:mRAC_LORAKS} can be minimized using existing algorithms for LORAKS optimization \cite{kim2018b, lobos2018navigator, haldar2014low, Haldar_2015}.  In particular, it is not hard to show that the solution to Eq.~\eqref{eq:mRAC_LORAKS} can be equivalently obtained by  solving:
\begin{equation}
\begin{split}
\left\{\hat{\mathbf{k}}^+,\hat{\mathbf{k}}^-\right\} & =  \argmin_{\left\{\mathbf{k}^+,\mathbf{k}^-\right\}} J_{(C-p)}\left(\begin{bmatrix} \mathcal{P}_C(\mathbf{k}^+) \\ \mathcal{P}_C(\mathbf{k}^-) \\ \sqrt{\eta}\mathcal{P}_C(\mathbf{k}^+_\text{acs}) \\ \sqrt{\eta}\mathcal{P}_C(\mathbf{k}^-_\text{acs})\end{bmatrix}\right) + \lambda J_r\left(\begin{bmatrix}
\mathcal{P}_S(\mathbf{k}^+)  &  \mathcal{P}_S(\mathbf{k}^-)  & \mathcal{P}_S(\mathbf{k}_{\text{acs}}^+)   & \mathcal{P}_S(\mathbf{k}_{\text{acs}}^-) 
\end{bmatrix}\right),
\end{split}\label{eq:4}
\end{equation}
where $J_{(C-p)}(\cdot)$ is the same  as $J_r(\cdot)$ but replacing the rank parameter $r$ with the rank parameter $(C-p)$, where $C$ is the number of columns of the LORAKS matrix formed by $\mathcal{P}_C(\cdot)$.  Equation~\eqref{eq:4} is convenient because it takes the same form as previous LORAKS optimization problems involving multiple $J_r(\cdot)$ terms \cite{haldar2014low}.   For this paper, we use a multiplicative half-quadratic majorize-minimize algorithm to minimize this objective function \cite{kim2018b}, which takes advantage of FFT-based matrix multiplications to improve computational complexity \cite{ongie2017}.

The RAC-LORAKS solution is obtained through the optimization of a nonconvex cost function.  As
such, the algorithm has the potential to converge to an undesirable 
local minimum.  For the results shown in this paper, we initialize RAC-LORAKS using a naive initialization with minimal processing cost as explained in the next section.  Other choices could potentially result in even higher performance, but are not considered here.

\section{METHODS}
\subsection{Datasets used for Evaluation}

As described below, we evaluated the characteristics of RAC-LORAKS using data from several different contexts.   All in vivo data were acquired under IRB-approved written informed consent. 

\subsubsection{Gradient-Echo  EPI Brain data }\label{sec:geepi}
 
 In one set of experiments, we acquired in vivo human brain data using a gradient-echo EPI sequence with  parameters that are somewhat similar to a BOLD fMRI experiment.  Data was acquired on a Siemens 3T Prisma Fit scanner using a standard 32-channel receiver array.  The data was acquired using FOV = 220 mm $\times$ 220 mm; matrix size = 128 $\times$ 128; slice thickness = 3 mm; and TR = 2 sec.  In one subject, data was acquired without acceleration ($R=1$) with TE = 47 msec.  From this same subject, data was also acquired for parallel imaging acceleration factors of $R=2,3,4$ with TE = 35 msec. In a second subject, data was acquired for parallel imaging acceleration factors of $R=5,6$ with TE = 35 msec.  In all cases, fully-sampled ACS data was acquired using the same interleaved $2R$-shot EPI prescan as used for DPG \cite{hoge2015dual}.
 
The previous datasets were acquired  with a conventional axial slice orientation. However, because Nyquist ghost problems tend to be more extreme with oblique acquisitions due to concomitant fields that can produce substantial nonlinear 2D phase differences between positive and negative readout polarities \cite{Xu_2010,hoge2015dual,reeder_1999,grieve_2002,chen_2011}, we also acquired an additional dataset with a double-oblique slice orientation from a third subject to test performance in a more challenging scenario. The slice orientation in this case is nonstandard and likely difficult to interpret for many readers, so we have depicted its position in Supporting Information Fig. S2. For this case the data was acquired with TR = 2.08 sec and  TE = 35 msec for parallel imaging acceleration factors  of $R = 1,2,3,4,5,6$.

\subsubsection{Diffusion-encoded  EPI Brain Data}
In another set of experiments, we acquired in vivo human brain data using a diffusion-encoded spin-echo EPI sequence.   Diffusion EPI data might be considered more challenging than the previous gradient-echo EPI data, due to the fact that diffusion MRI data usually suffers from random image-to-image phase variations, and can also have lower SNR than gradient-echo EPI. In addition, the rapid switching of strong diffusion gradients can introduce substantial additional eddy current effects that can cause systematic differences between the ACS data and the diffusion EPI data if they are acquired with different diffusion gradient settings \cite{le2006}.  

A first diffusion dataset was acquired on a Siemens 3T Prisma Fit scanner using a standard 32-channel receiver array.  For the sake of computational complexity, this data was subsequently reduced to 16 channels using standard coil-compression techniques.  The data was acquired using FOV = 220 mm $\times$ 220 mm; matrix size = 220 $\times$ 220; slice thickness = 5 mm; TR = 2.8 sec; TE = 63 msec; b-values of 0 sec/mm$^2$ and 1000 sec/mm$^2$; 6 diffusion encoding directions; parallel imaging acceleration factor $R=3$; and 6/8ths partial Fourier sampling.  ACS data was acquired using the same interleaved $2R$-shot EPI prescan as used for DPG \cite{hoge2015dual}, except that the data was acquired with lower resolution along the phase encoding dimension (i.e., we only acquired 45 phase-encoding lines for the ACS data). Due to the random phase variations associated with diffusion encoding gradients, the ACS data was acquired without diffusion weighting, which means that the ACS data has very different contrast characteristics from the EPI data. To show results across a broader range of acceleration factors, a second set of acquisitions  was performed with $R=2,3,4,5$.  Other parameters were identical  to the previous case, except for matrix size = 110 $\times$ 110; slice thickness = 2 mm; TR = 11.4 sec; TE = 73 msec; and fully-sampled ACS data.

\subsubsection{Cardiac  EPI Data}
In a third set of experiments, we acquired in vivo human cardiac data during diastole using a spin-echo EPI sequence with parameters that are typical for a myocardial arterial spin labeling experiment \cite{javed2020}.  Data was acquired on a GE 3T Signa HDx scanner with an 8-channel cardiac coil.  The acquisition used FOV = 280 mm $\times$ 140 mm; matrix size = 128 $\times$ 64; slice thickness = 10 mm; TR = 55 msec; TE = 32.9 msec; velocity cutoff = 5 cm/s; no parallel imaging acceleration ($R=1$); and 5/8ths partial Fourier sampling.   ACS data was acquired using the same interleaved $2R$-shot EPI prescan as used for DPG \cite{hoge2015dual}, but with 5/8ths partial Fourier sampling.  Data was acquired with a double-oblique slice orientation to achieve a  mid-short axis view.

\subsection{Simulations}

In addition to in vivo data, the different methods were also evaluated using simulations where a gold standard was present.   To form a gold standard with realistic EPI characteristics, we took two in vivo gradient-echo EPI brain datasets (as described in Section~\ref{sec:geepi}) with axial slice orientation and $R=1$ from the same scan session, and reconstructed them both using SENSE. Each gradient polarity was reconstructed separately, providing a realistic representation of typical interpolarity image differences. This procedure provides two sets of fully-sampled multi-channel dual-polarity gold standard images. One of these sets was used for ACS data, while the other was undersampled (including parallel imaging acceleration, along with interleaving the data from positive and negative gradient polarities) to simulate EPI data.  These datasets were acquired roughly 5 minutes apart, allowing time for mismatches to evolve. Since ghost correction is frequently more difficult for EPI datasets with 2D nonlinear phase differences between the two polarities, we applied an additional 2D nonlinear phase pattern to make the problem more challenging. This additional phase difference was designed to be roughly $3\times$-larger than we observed in the real data from Fig. 2.   

In a first set of simulations, to mimic the situation where a localized image feature is different between the ACS data and EPI data (e.g., as may happen in a dynamic experiment), we added a Gaussian-shaped additive image hyperintensity to the EPI data that we did not add to the ACS data.  The hyperintensity was designed to follow both the coil sensitivity maps (obtained by applying ESPIRiT \cite{uecker2014}) and the phase characteristics of the original data.  We also performed simulations with these two datasets interchanged, i.e., with the hyperintensity in the ACS data but not in the EPI data.

In another set of simulations, to mimic the situation where the ACS data and EPI data have very different contrasts, we inverted the magnitude image for the ACS data to create an image with different contrast \cite{bhushan2015co}, while still following the coil sensitivity maps and the phase characteristics of the original data.

In addition to performing multi-channel simulations, we also performed a simulation in a very challenging single-channel setting.  For this, single-channel data was obtained by a linear combination of the multi-channel data \cite{buehrer2009virtual}.  Single-channel ghost correction is a difficult setting where only a few previous methods have had any success \cite{lobos2018navigator,Lee_2016}.  This case is hard because even with unaccelerated data ($R=1$), each polarity has an effective acceleration factor of $R=2$ when the data for each readout gradient polarity is separated, and it can be difficult to reconstruct $R=2$ data without multi-channel information.

 The fully-sampled ACS and EPI datasets used for all three simulations are illustrated in Supporting Information Fig.~S3.

\subsection{Data Processing}

RAC-LORAKS was applied to perform reconstruction and ghost correction on these datasets.  For comparison against existing methods, the datasets were also reconstructed using the previous AC-LORAKS ghost-correction method \cite{lobos2018navigator}, DPG \cite{hoge2015dual}, and MUSSELS \cite{Mani_2016}. 

For some of the datasets we consider, the ACS data may be incomplete due to low-resolution ACS acquisition (i.e., the first brain EPI diffusion data) or partial Fourier ACS acquisition (i.e., the cardiac EPI data).  In such cases, we modify RAC-LORAKS to consider the fully sampled ACS data vectors $\mathbf{k}^{+}_\mathrm{acs}$ and $\mathbf{k}^{-}_\mathrm{acs}$ as additional variables to be optimized in Eq.~\eqref{eq:4}, subject to ACS data consistency constraints.

For RAC-LORAKS and AC-LORAKS, the regularization parameters $\lambda$ and $\eta$ were selected manually based on subjective visual inspection reconstruction quality and ghost-reduction performance for in vivo data, and to minimize quantitative error measures for simulated data.  The rank-related parameters $p$ and $r$ were selected based on the singular value characteristics of LORAKS matrices formed from the ACS data.  The rank parameters were set based on the points at which the singular value curves begin to flatten out, which is a common rank estimation technique for noisy matrices.   This decision was made manually (based on visual inspection) for the results shown later in the paper, although fully automatic approaches would also be viable.

DPG is a ghost correction method that treats different gradient polarities like different coils in a parallel imaging experiment, and uses a dual GRAPPA kernel estimated from ACS data for image reconstruction  \cite{hoge2015dual}.  In order to use DPG for the initialization of RAC-LORAKS, we have adapted DPG to output two sets of images  (with calibration based on the raw uncorrected multi-channel ACS data), one for the original $\mathbf{k}^+$ data and one for the original $\mathbf{k}^-$ data. This is different than the original DPG implementation, which applies ACS pre-processing to try and correct for errors in the ACS data, and then directly fuses information from the two polarities together into a single virtual ``hybrid'' output \cite{hoge2015dual}.  This hybrid output can have different magnitude and phase characteristics than the original $\mathbf{k}^+$ and $\mathbf{k}^-$ data, so is not useful as an initialization for RAC-LORAKS.   We refer to our adapted version as modified DPG (mDPG) from now on.  In some cases, we also compare against the original version of DPG (including the original ACS pre-processing procedure to correct for shot-to-shot variations in the ACS data \cite{hoge2015dual}), although note that such comparisons are necessarily qualitative, since the magnitude and phase characteristics of the hybrid  output DPG images do not match the images generated using other methods. 

MUSSELS is a structured low-rank matrix recovery method that uses SENSE-type parallel imaging constraints together with nuclear norm regularization to impose low-rank constraints \cite{Mani_2016}.  While MUSSELS was originally developed for multi-shot EPI ghost correction, it can apply equally well to the ghost correction problem associated with different gradient polarities.  Sensitivity maps for MUSSELS were estimated by applying ESPIRiT \cite{uecker2014} to the same ACS data used for the other methods.  The regularization parameter for MUSSELS was selected manually based on subjective visual inspection of reconstruction quality and ghost-reduction performance in the case of in vivo data, or to minimize quantitative error measures for simulated data.

Note that DPG and MUSSELS were both developed for the multi-channel setting.  We can adapt DPG  to the single-channel setting in straightforward ways \cite{lobos2018navigator}, and we apply this adaptation to the single-channel simulated data.  We did not adapt MUSSELS to the single-channel case.  Note that the SENSE-based constraints used by MUSSELS would reduce to a simple spatial-domain support constraint in the single-channel case, which is not strong enough to yield good performance results.

For all methods, results were visualized by using a standard square-root sum-of-squares technique to combine the images from different coils and different gradient polarities into a single image.  Results from in vivo experiments were evaluated qualitatively, since a gold standard reference was not available in these cases.

Simulation results were evaluated quantitatively  using the normalized root mean-squared error (NRMSE):
\begin{equation}
 \mathrm{NRMSE} \triangleq \frac{\sqrt{\left\|\hat{\mathbf{k}}^+ - \mathbf{k}^+_{\mathrm{gold}}\right\|_2^2+\left\|\hat{\mathbf{k}}^- - \mathbf{k}^-_{\mathrm{gold}}\right\|_2^2}}{\sqrt{\left\| \mathbf{k}^+_{\mathrm{gold}}\right\|_2^2+\left\| \mathbf{k}^-_{\mathrm{gold}}\right\|_2^2}},
\end{equation}
where $\mathbf{k}^+_{\mathrm{gold}}$ and $\mathbf{k}^-_{\mathrm{gold}}$ are respectively the gold standard values for the positive and negative gradient polarities. We also plotted Fourier Error Spectrum Plots (ESPs) to gain further insight into how the errors were distributed across different spatial resolutions scales \cite{kim2018fourier}.  An ESP is designed to reveal the spectral characteristics of the error, and for example, can discriminate between methods that make more errors in the low-resolution features of an image versus methods that make more errors in high-resolution features.

\section{RESULTS}
Figure~1 shows ACS data and reconstruction results from the in vivo gradient-echo EPI brain data with an axial slice orientation.    The ACS data in this case does not have strong artifacts, although close inspection does reveal that ACS ghost artifacts are present. This can be further appreciated in Supporting Information Fig.~S4 where the same images are shown with amplified image intensity to highlight ghost characteristics in the image background.  As can be seen, all ghost correction methods work well at smaller acceleration factors, although performance begins to degrade at larger acceleration factors.  We observe that, compared to other methods, the visual quality of the MUSSELS reconstruction seems to degrade most rapidly as a function of acceleration factor,  which is consistent with previous observations \cite{lobos2018navigator}.  The mDPG method had qualitatively better performance than MUSSELS in this case.  However, a close inspection of the images reveals that the mDPG results are not entirely ghost-free even for the unaccelerated ($R=1$) case.  This may be expected due to the artifacts and mismatches that are present in the ACS data.  Although mDPG does not attempt to correct the ACS artifacts, it should be noted that the original DPG method does try to correct them through pre-processing.  Results showing the qualitative performance of the original DPG method are shown in Supporting Information Fig.~S5, where we observe that the ghost artifacts still exist, though as expected, are less prominent than were observed for mDPG.  In spite of the ACS artifacts, the AC-LORAKS reconstruction still has good performance at low acceleration factors  and does a good job of suppressing ghosts in the background regions of the image at all acceleration factors, although exhibits substantial degradation in image quality at the highest acceleration factors (with artifacts similar to those observed for highly-accelerated parallel imaging reconstructions).  However, the RAC-LORAKS reconstruction appears to have much higher quality than the other methods, even at very high acceleration factors like $R=6$.  (Note that when $R=6$, the effective acceleration factor is $R=12$ when each readout gradient polarity is considered separately.  This leads to a highly ill-posed inverse problem).

Figure~2 shows results from the in vivo gradient-echo EPI brain data with a  double-oblique slice orientation.  This case is more challenging than the previous one due to the complicated nonlinear 2D spatial phase differences we observed between data acquired with positive and negative polarities (as visualized in the last column of Fig.~2), the proximity to air-tissue interfaces that result in substantial magnetic field inhomogeneity effects, as well as more substantial ghosting artifacts present in the ACS data. Note that the ACS data corresponding to the $R=5$ case is particularly corrupted, which can be attributed to the unpredictable shot-to-shot variations that frequently occur in these kinds of multi-shot acquisitions. Despite the more extreme scenario, the different ghost reconstruction methods have similar characteristics to those observed in the previous case, with RAC-LORAKS appearing to demonstrate the cleanest overall results. 

 Figure~3 shows reconstruction results from the first set of multi-channel simulations (with similar contrast between ACS and EPI data, but with a hyperintensity added to the EPI data). Quantitative NRMSE values are reported in Table~1 with corresponding ESPs shown in Fig.~4. Qualitatively, the results from Fig.~3 have similar characteristics to the results observed with in vivo data.  Notably, RAC-LORAKS is able to consistently reconstruct a high-quality image that bears close resemblance to the gold standard image, while methods like mDPG and AC-LORAKS have artifacts due to the small mismatches between the ACS and EPI data.   The visual assessment of reconstruction quality matches well with the quantitative NRMSE assessment shown in Table~1.  AC-LORAKS and RAC-LORAKS have a similar performance at $R = 1$ and $2$, with RAC-LORAKS having the best performance at high acceleration factors.

Reconstructions were also performed using the original DPG formulation as shown in Supporting Information Fig.~S6.  In this case, DPG has similar ghost artifacts to mDPG, which is expected because there are no artifacts in the ACS data, while there is a problematic mismatch between the ACS data and the EPI data that neither DPG nor mDPG address.   Notably, for both DPG and mDPG, we observe aliasing artifacts that seem to be associated with the hyperintensity that was present in the EPI data but was  not in the ACS data.

The ESP plots in Fig.~4 enable a more nuanced analysis.  These results suggest that RAC-LORAKS has good (i.e., among the best, even if it is not always the best) performance at all spatial frequencies, meaning that it is good at reconstructing image features across the whole range of resolution scales. 

 Supporting Information Fig. S7 shows a similar simulation result to that shown in Fig. 3, with the main difference being that the previous EPI images (with the hyperintensity) were used as ACS data and the previous ACS images (without the  hyperintensity) were used to generate EPI data.  Consistent with the previous case, we observe good performance for RAC-LORAKS, and do not observe the features of the hyperintensity being erroneously transferred into the RAC-LORAKS reconstruction results. 
 
 Supporting Information Fig.~S8 and Supporting Information Table~S1 show simulation results for the case where the ACS data has an even more substantial contrast difference (i.e. inverted contrast)  with respect to the EPI data.  For this case we observe a degradation in performance for all methods compared to the previous cases, although RAC-LORAKS still showed the best overall qualitative and quantitative performance.  This result suggests that RAC-LORAKS may have better performance when the contrast is similar between the ACS and EPI data, although can still provide benefits when the contrast difference is substantial.

Figure~5 shows reconstruction results from the single-channel simulation, with quantitative NRMSE values reported in Supporting Information Table~S2. While previous work \cite{lobos2018navigator} reported that mDPG and AC-LORAKS can be reasonably successful for single-channel data with $R=1$ when the ACS data is pristine, our new results demonstrate that this performance can be sensitive to the quality of the ACS data.    In particular, we observe strong ghost artifacts for both of these methods, even though we do observe that the AC-LORAKS reconstruction has successfully suppressed ghost artifacts in the image background (outside of the support of the true image).  In contrast, RAC-LORAKS is substantially more successful for $R=1$.  Notably, RAC-LORAKS also performed well for the even more challenging $R=2$ case, unlike the other methods.  For reference, note that even with high-quality ACS data, the previous AC-LORAKS method did not yield good results with similar single-channel $R=2$  data  \cite{lobos2018navigator}.

Figure~6 shows reconstruction results from the first set of in vivo diffusion EPI brain data, including a $10\times$ intensity amplification to highlight the ghost characteristics.  As can be seen, the ACS data has ghost artifacts in all cases, and both MUSSELS and mDPG reconstructions also exhibit unsuppressed ghosting artifacts.  On the other hand, both AC-LORAKS and RAC-LORAKS are relatively ghost-free in this example and have only minor differences from one another (it might be argued that the RAC-LORAKS result has a slightly less-noisy appearance than the AC-LORAKS result, but if so, this difference is very subtle).  While this result does not demonstrate an obvious advantage for RAC-LORAKS over AC-LORAKS, it should be observed that this diffusion result is at least consistent with the previous gradient-echo EPI data results, in which we also did not observe a substantial difference between RAC-LORAKS and AC-LORAKS when $R=3$.  In addition, this case involves a very substantial contrast difference between the ACS data and the EPI data.  This difference does not appear to have adversely affected the performance characteristics of these methods in substantial ways. 

Figure~7 shows reconstruction results from the second set of in vivo diffusion EPI brain acquisitions (with different acceleration factors), with zoom-ins shown in Supporting Information Fig. S9 for improved visibility. Consistent with the results shown for the gradient-echo EPI data in Fig.~1, we observe that all methods perform well for low acceleration factors. As the acceleration factor increases, the performance of each method degrades,  with RAC-LORAKS showing a lower qualitative degradation in comparison to the other methods at the very high acceleration factors $R =4,5$. Note that at high acceleration factors (e.g., $R=4,5$) the reconstruction quality for RAC-LORAKS is not quite as good as for the gradient-echo EPI dataset shown in Fig.~1.  We believe that this should be expected, since as mentioned before, diffusion EPI data can be considered more challenging than the gradient-echo EPI data  due to SNR issues, eddy current effects, motion-induced phase effects, and contrast mismatches between the ACS and EPI datasets.

Finally, Fig.~8 shows results from the in vivo cardiac EPI data.  While this data was not accelerated ($R=1$), this case is challenging because of the double-oblique slice orientation as well as the substantial artifacts present in the ACS data resulting from cardiac motion-induced shot-to-shot variations.  In addition, this case can also be challenging for SENSE-based methods (like MUSSELS), due to the use of a small FOV with aliasing.  When aliasing is present within the FOV, it violates the standard SENSE modeling assumption of one sensitivity map value per spatial location, which generally leads to artifacts if not properly accounted for.    The results demonstrate that both MUSSELS and mDPG have substantial residual ghosting artifacts, which might not be surprising given the high degree of corruption that is present in the ACS data.  On the other hand, both AC-LORAKS and RAC-LORAKS are more successful at suppressing the ghosts.  Without a gold standard reference, it is hard to establish definitively whether AC-LORAKS or RAC-LORAKS is better in this example, although we believe that the RAC-LORAKS result demonstrates slightly less ghosting  than AC-LORAKS, particularly on the left side of the image where the ACS data and mDPG both have particularly strong ghost artifacts.  

\section{DISCUSSION}

The results in the previous section demonstrated that, in the presence of imperfect ACS data, RAC-LORAKS frequently offers similar or better performance to the previous AC-LORAKS ghost correction method that it generalizes, while both of these methods perform substantially better than methods like MUSSELS or DPG.  We also observed that RAC-LORAKS appears to have the biggest advantage over AC-LORAKS in scenarios where the parallel imaging acceleration factor was high. For these cases, we observed that RAC-LORAKS was able to mitigate ghost artifacts both inside and outside the support of the original image, while AC-LORAKS was only able to mitigate ghost artifacts outside the support but not inside. This advantage for RAC-LORAKS is likely the result of its improved robustness to ACS errors combined with the multi-contrast linear predictability constraints which help to make the reconstruction problem less ill-posed. However, it should be noted that  RAC-LORAKS has one more regularization parameter  than AC-LORAKS (i.e., $\eta$, which controls the level of trust placed in the information from the ACS data). In our experience, manual tuning of this parameter is not hard (i.e., we always started from the small value  $\eta=10^{-3}$, and frequently did not have to modify this value to achieve satisfying results).  The method would be easier to use if the selection of $\eta$ were automated. 

Both RAC-LORAKS and AC-LORAKS also depend on the choice of rank parameters, and as described previously, the results shown in this work made a heuristic choice based on the empirical rank characteristics of the ACS data. Even though the low-rank characteristics of the structured matrices might vary between the ACS data and the acquired EPI data due to systematic phenomena (e.g., thermal noise, subject motion, respiration, artifacts in the ACS data, etc.), we have not observed major problems associated with inappropriate rank selection in our empirical results.  This might be expected, based on the observation that LORAKS reconstruction results are frequently not very sensitive to small variations in the rank parameter \cite{haldar2014low,Haldar_2015}.  Nevertheless, the development of improved automatic RAC-LORAKS parameter selection methods would be an interesting topic for future work.

Although RAC-LORAKS offers good performance, it should be noted that our current implementation of RAC-LORAKS can be more computationally expensive than existing methods.  For example, for the results shown with $R=1$ in Fig.~1, RAC-LORAKS used $\approx 45$ min of reconstruction time, while MUSSELS, mDPG, and AC-LORAKS respectively used $\approx 15$ min, $\approx 2$ min, and $\approx 100$ min.  All methods were implemented in MATLAB on a standard desktop computer with an Intel Xeon E5-1603 2.8 GHz quad core CPU processor and 32GB of RAM. While this relatively long computation time may be a concern, it should be noted that we are reporting the results of a simple proof-of-principle implementation, and we did not spend much time to optimize the computational efficiency of this approach.  We believe that major improvements may be possible by leveraging better computational hardware, smarter algorithms, and more efficient implementations.  Given the reconstruction performance offered by RAC-LORAKS, we believe that improving its computational performance is a promising topic for future research. However, RAC-LORAKS is notably faster than AC-LORAKS, and it appears that this speed difference results from the fact that RAC-LORAKS has consistently faster convergence than AC-LORAKS in this setting.  The reason for this faster convergence is unclear at this stage, although we believe that a detailed analysis of convergence characteristics is beyond the scope of the present paper.

While this paper focused on EPI ghost correction for standard single-slice excitation, we believe that the extension of these ideas to simultaneous multi-slice EPI acquisitions (similar to Refs.~\cite{Lyu_2018, Liu_2019,mani2019sms_conf,bilgic2019}) is a very promising research direction.

Finally, although the techniques we developed in this work were described and evaluated in the context of EPI ghost correction, we believe that the overall approach is likely to be useful across a wide range of parallel imaging applications, particularly those for which the measured ACS data is not adequate to resolve all of the reconstruction artifacts.  Specifically, we believe that the key  principles employed by RAC-LORAKS (i.e., using structured low-rank matrix methods to avoid placing complete trust in the accuracy of ACS data, and leveraging ACS data to provide additional information in a multi-contrast framework) are both novel ideas that are applicable to arbitrary image reconstructions involving ACS data, and are not exclusive to ghost correction settings. In addition, we are encouraged by the high-quality reconstruction results that RAC-LORAKS produces even in very highly-accelerated scans. These results suggest to us that there may be value in exploring the usefulness of RAC-LORAKS to other parallel imaging experiments in future work.

\section{CONCLUSIONS}

This work proposed and evaluated RAC-LORAKS, a new structured low-rank matrix method for EPI ghost correction  that integrates multiple  constraints (including parallel imaging constraints, support constraints, phase constraints, and inter-image linear predictability constraints) to not only mitigate artifacts resulting from imperfect ACS data and Nyquist ghosts, but also accounting for partial Fourier acquisition and reducing parallel imaging artifacts and noise in an integrated fashion.  RAC-LORAKS uses ACS data and k-space domain linear predictive modeling to stabilize the solution of the ill-posed inverse problem, and was observed to offer advantages relative to state-of-the-art ghost correction methods like AC-LORAKS, DPG, and MUSSELS.

\section*{ACKNOWLEDGMENTS}
This work was supported in part by NSF CAREER award CCF-1350563 and NIH grants R01-MH116173, R21-EB022951, R01-NS089212, R01-NS074980, and NIH R01-HL130494. Computation for some of the work described in this paper was supported by the University of Southern California's Center for High-Performance Computing (http://hpcc.usc.edu/).

\bibliographystyle{MRMbib}
\bibliography{./bibliography}

\providecommand{\noopsort}[1]{}
\begin{thebibliography}{10}

\bibitem{stehling1991}
Stelhing MK, Turner R, Mansfield P. Echo-planar imaging: Magnetic resonance
  imaging in a fraction of a second. Science 1991;\hspace{0pt}254:43--50.

\bibitem{bernstein2004}
Bernstein MA, King KF, Zhou XJ. \emph{Handbook of MRI Pulse Sequences}.
  Burlington: Elsevier Academic Press. 2004.

\bibitem{Bruder_1992}
Bruder H, Fischer H, Reinfelder HE, Schmitt F. Image reconstruction for echo
  planar imaging with nonequidistant k-space sampling. Magn{} Reson{} Med{}
  1992;\hspace{0pt}23:311--323.

\bibitem{Hu_1996}
Hu X, Le TH. Artifact reduction in {EPI} with phase-encoded reference scan.
  Magn{} Reson{} Med{} 1996;\hspace{0pt}36:166--171.

\bibitem{Buonocore_1997}
Buonocore MH, Gao L. Ghost artifact reduction for echo planar imaging using
  image phase correction. Magn{} Reson{} Med{} 1997;\hspace{0pt}38:89--100.

\bibitem{Chen_2004}
Chen Nk, Wyrwicz AM. Removal of {EPI} {N}yquist ghost artifacts with
  two-dimensional phase correction. Magn{} Reson{} Med{}
  2004;\hspace{0pt}51:1247--1253.

\bibitem{skare2006fast}
Skare S, Clayton D, Newbould R, Moseley M, Bammer R. A fast and robust minimum
  entropy based non-interactive {N}yquist ghost correction algorithm. In:
  \emph{Proc. Int. Soc. Magn. Reson. Med.}. 2006;\hspace{0pt} p. 2349.

\bibitem{xiang2007correction}
Xiang QS, Ye FQ. Correction for geometric distortion and {N}/2 ghosting in
  {EPI} by phase labeling for additional coordinate encoding ({PLACE}). Magn{}
  Reson{} Med{} 2007;\hspace{0pt}57:731--741.

\bibitem{hoge2010robust}
Hoge WS, Tan H, Kraft RA. Robust {EPI} {N}yquist ghost elimination via spatial
  and temporal encoding. Magn{} Reson{} Med{} 2010;\hspace{0pt}64:1781--1791.

\bibitem{Xu_2010}
Xu D, King KF, Zur Y, Hinks RS. Robust 2{D} phase correction for echo planar
  imaging under a tight field-of-view. Magn{} Reson{} Med{}
  2010;\hspace{0pt}64:1800--1813.

\bibitem{hoge2015dual}
Hoge WS, Polimeni JR. Dual-polarity {GRAPPA} for simultaneous reconstruction
  and ghost correction of echo planar imaging data. Magn{} Reson{} Med{}
  2016;\hspace{0pt}76:32--44.

\bibitem{chang2016}
Chang HC, Chen NK. Joint correction of {N}yquist artifact and minuscule
  motion-induced aliasing artifact in interleaved diffusion weighted {EPI} data
  using a composite two-dimensional phase correction procedure. Magn Reson Imag
  2016;\hspace{0pt}34:974--979.

\bibitem{Yarach_2017}
Yarach U, In MH, Chatnuntawech I, Bilgic B, Godenschweger F, Mattern H, Sciarra
  A, Speck O. Model-based iterative reconstruction for single-shot {EPI} at
  7{T}. Magn{} Reson{} Med{} 2017;\hspace{0pt}78:2250--2264.

\bibitem{Ianni_2017}
Ianni JD, Welch EB, Grissom WA. Ghost reduction in echo-planar imaging by joint
  reconstruction of images and line-to-line delays and phase errors. Magn{}
  Reson{} Med{} 2018;\hspace{0pt}79:3114--3121.

\bibitem{Xie_2017}
Xie VB, Lyu M, Liu Y, Feng Y, Wu EX. Robust {EPI} {N}yquist ghost removal by
  incorporating phase error correction with sensitivity encoding
  ({PEC}-{SENSE}). Magn{} Reson{} Med{} 2018;\hspace{0pt}79:943--951.

\bibitem{Lee_2016}
Lee J, Jin KH, Ye JC. Reference-free {EPI} {N}yquist ghost correction using
  annihilating filter-based low rank {H}ankel matrix for {K}-space
  interpolation. Magn{} Reson{} Med{} 2016;\hspace{0pt}76:1775--1789.

\bibitem{Mani_2016}
Mani M, Jacob M, Kelley D, Magnotta V. Multi-shot sensitivity-encoded diffusion
  data recovery using structured low-rank matrix completion ({MUSSELS}). Magn{}
  Reson{} Med{} 2017;\hspace{0pt}78:494--507.

\bibitem{lobos2018navigator}
Lobos RA, Kim TH, Hoge WS, Haldar JP. Navigator-free {EPI} ghost correction
  with structured low-rank matrix models: {N}ew theory and methods. IEEE Trans
  Med Imag 2018;\hspace{0pt}37:2390--2402.

\bibitem{Lyu_2018}
Lyu M, Barth M, Xie VB, Liu Y, Ma X, Feng Y, Wu EX. Robust {SENSE}
  reconstruction of simultaneous multislice {EPI} with low-rank enhanced coil
  sensitivity calibration and slice-dependent 2{D} {N}yquist ghost correction.
  Magn{} Reson{} Med{} 2018;\hspace{0pt}80:1376--1390.

\bibitem{Liu_2019}
Liu Y, Lyu M, Barth M, Yi Z, Leong AT, Chen F, Feng Y, Wu EX. {PEC-GRAPPA}
  reconstruction of simultaneous multislice {EPI} with slice-dependent 2{D}
  {N}yquist ghost correction. Magn{} Reson{} Med{}
  2019;\hspace{0pt}81:1924--1934.

\bibitem{haldar2015autocalibrated}
Haldar JP. Autocalibrated {LORAKS} for fast constrained {MRI} reconstruction.
  In: \emph{Proc. IEEE Int. Symp. Biomed. Imag.}. 2015;\hspace{0pt} pp.
  910--913.

\bibitem{griswold2002}
Griswold MA, Jakob PM, Heidemann RM, Nittka M, Jellus V, Wang J, Kiefer B,
  Haase A. Generalized autocalibrating partially parallel acquisitions
  ({GRAPPA}). Magn{} Reson{} Med{} 2002;\hspace{0pt}47:1202--1210.

\bibitem{lustig2010}
Lustig M, Pauly JM. {SPIRiT}: Iterative self-consistent parallel imaging
  reconstruction from arbitary \emph{k}-space. Magn{} Reson{} Med{}
  2010;\hspace{0pt}65:457--471.

\bibitem{zhang2011}
Zhang J, Liu C, Moseley ME. Parallel reconstruction using null operations.
  Magn{} Reson{} Med{} 2011;\hspace{0pt}66:1241--1253.

\bibitem{ying2007}
Ying L, Sheng J. Joint image reconstruction and sensitivity estimation in
  {SENSE} ({JSENSE}). Magn{} Reson{} Med{} 2007;\hspace{0pt}57:1196--1202.

\bibitem{sheltraw2012}
Sheltraw D, Inglis B, Deshpande V, Trumpis M. Simultaneous reduction of two
  common autocalibration errors {GRAPPA} {EPI} time series data. Preprint
  2012;\hspace{0pt}{arXiv:1208.0972}.

\bibitem{baron2016}
Baron CA, Beaulieu C. Motion robust {GRAPPA} for echo-planar imaging. Magn{}
  Reson{} Med{} 2016;\hspace{0pt}75:1166--1174.

\bibitem{talagala2016}
Talagala SL, Sarlls JE, Liu S, Inati SJ. Improvement of temporal
  signal-to-noise ratio of {GRAPPA} accelerated echo planar imaging using a
  {FLASH} based calibration scan. Magn{} Reson{} Med{}
  2016;\hspace{0pt}75:2362--2371.

\bibitem{holme2019}
Holme HCM, Rosenzweig S, Ong F, Wilke RN, Lustig M, Uecker M. {ENLIVE}: An
  efficient nonlinear method for calibrationless and robust parallel imaging.
  Sci Rep 2019;\hspace{0pt}9:1--13.

\bibitem{bilgic2018improving}
Bilgic B, Kim TH, Liao C, Manhard MK, Wald LL, Haldar JP, Setsompop K.
  Improving parallel imaging by jointly reconstructing multi-contrast data.
  Magn{} Reson{} Med{} 2018;\hspace{0pt}80:619--632.

\bibitem{lobos2018robust}
Lobos RA, Javed A, Nayak KS, Hoge WS, Haldar JP. Robust autocalibrated {LORAKS}
  for {EPI} ghost correction. In: \emph{Proc. IEEE Int. Symp. Biomed. Imag.}.
  2018;\hspace{0pt} pp. 663--666.

\bibitem{lobos2018robust_ismrm}
Lobos RA, Javed A, Nayak KS, Hoge WS, Haldar JP. Robust autocalibrated {LORAKS}
  for improved {EPI} ghost correction with structured low-rank matrix models.
  In: \emph{Proc. Int. Soc. Magn. Reson. Med.}. 2018;\hspace{0pt} p. 3533.

\bibitem{liang1989}
Liang ZP, Haacke EM, Thomas CW. High-resolution inversion of finite {F}ourier
  transform data through a localised polynomial approximation. Inverse Probl
  1989;\hspace{0pt}5:831--847.

\bibitem{uecker2014}
Uecker M, Lai P, Murphy MJ, Virtue P, Elad M, Pauly JM, Vasanawala SS, Lustig
  M. {ESPIRiT} -- an eigenvalue approach to autocalibrating parallel {MRI}:
  Where {SENSE} meets {GRAPPA}. Magn{} Reson{} Med{}
  2014;\hspace{0pt}71:990--1001.

\bibitem{shin2014}
Shin PJ, Larson PEZ, Ohliger MA, Elad M, Pauly JM, Vigneron DB, Lustig M.
  Calibrationless parallel imaging reconstruction based on structured low-rank
  matrix completion. Magn{} Reson{} Med{} 2014;\hspace{0pt}72:959--970.

\bibitem{haldar2014low}
Haldar JP. Low-{R}ank {M}odeling of {L}ocal-{S}pace {N}eighborhoods ({LORAKS})
  for {C}onstrained {MRI}. IEEE Trans Med Imag 2014;\hspace{0pt}33:668--681.

\bibitem{Haldar_2015}
Haldar JP, Zhuo J. P-{LORAKS}: Low-rank modeling of local k-space neighborhoods
  with parallel imaging data. Magn{} Reson{} Med{}
  2015;\hspace{0pt}75:1499--1514.

\bibitem{ongie2016}
Ongie G, Jacob M. Off-the-grid recovery of piecewise constant images from few
  {F}ourier samples. SIAM J Imaging Sci 2016;\hspace{0pt}9:1004--1041.

\bibitem{jin2015a}
Jin KH, Lee D, Ye JC. A general framework for compressed sensing and parallel
  {MRI} using annihilating filter based low-rank {H}ankel matrix. IEEE Trans
  Comput Imaging 2016;\hspace{0pt}2:480--495.

\bibitem{haldar2019linear}
Haldar JP, Setsompop K. Linear predictability in magnetic resonance imaging
  reconstruction: Leveraging shift-invariant {F}ourier structure for faster and
  better imaging. IEEE Signal Process Mag 2020;\hspace{0pt}37:69--82.

\bibitem{Pruessmann_1999}
Pruessmann KP, Weiger M, Scheidegger MB, Boesiger P. {SENSE}: Sensitivity
  encoding for fast {MRI}. Magn{} Reson{} Med{} 1999;\hspace{0pt}42:952--962.

\bibitem{polimeni2016reducing}
Polimeni JR, Bhat H, Witzel T, Benner T, Feiweier T, Inati SJ, Renvall V,
  Heberlein K, Wald LL. Reducing sensitivity losses due to respiration and
  motion in accelerated echo planar imaging by reordering the autocalibration
  data acquisition. Magn{} Reson{} Med{} 2016;\hspace{0pt}75:665--679.

\bibitem{hoge2018dual}
Hoge WS, Setsompop K, Polimeni JR. Dual-polarity slice-{GRAPPA} for concurrent
  ghost correction and slice separation in simultaneous multi-slice {EPI}.
  Magn{} Reson{} Med{} 2018;\hspace{0pt}80:1364--1375.

\bibitem{liao2019phase}
Liao C, Manhard MK, Bilgic B, Tian Q, Fan Q, Han S, Wang F, Park DJ, Witzel T,
  Zhong J, \emph{et~al.}. Phase-matched virtual coil reconstruction for highly
  accelerated diffusion echo-planar imaging. NeuroImage
  2019;\hspace{0pt}194:291--302.

\bibitem{kim2018b}
Kim TH, Haldar JP. {LORAKS} software version 2.0: Faster implementation and
  enhanced capabilities. Technical Report USC-SIPI-443. University of Southern
  California. Los Angeles, CA. 2018.

\bibitem{ongie2017}
Ongie G, Jacob M. A fast algorithm for convolutional structured low-rank matrix
  recovery. IEEE Trans Comput Imaging 2017;\hspace{0pt}3:535--550.

\bibitem{reeder_1999}
Reeder SB, Atalar E, Faranesh AZ, McVeigh ER. Referenceless interleaved
  echo-planar imaging. Magn{} Reson{} Med{} 1999;\hspace{0pt}41:87--94.

\bibitem{grieve_2002}
Grieve SM, Blamire AM, Styles P. Elimination of {N}yquist ghosting caused by
  read-out to phase-encode gradient cross-terms in {EPI}. Magn{} Reson{} Med{}
  2002;\hspace{0pt}47:337--343.

\bibitem{chen_2011}
Chen NK, Avram AV, Song AW. Two-dimensional phase cycled reconstruction for
  inherent correction of echo-planar imaging {N}yquist artifacts. Magn{}
  Reson{} Med{} 2011;\hspace{0pt}66:1057--1066.

\bibitem{le2006}
Le~Bihan D, Poupon C, Amadon A, Lethimonnier F. Artifacts and pitfalls in
  diffusion {MRI}. J Magn Reson Imag 2006;\hspace{0pt}24:478--488.

\bibitem{javed2020}
Javed A, Nayak KS. Single-shot {EPI} for {ASL}-{CMR}. Magn{} Reson{} Med{}
  2020;\hspace{0pt}84:738--750.

\bibitem{bhushan2015co}
Bhushan C, Haldar JP, Choi S, Joshi AA, Shattuck DW, Leahy RM. Co-registration
  and distortion correction of diffusion and anatomical images based on inverse
  contrast normalization. Neuroimage 2015;\hspace{0pt}115:269--280.

\bibitem{buehrer2009virtual}
Buehrer M, Boesiger P, Kozerke S. Virtual body coil calibration for
  phased-array imaging. In: \emph{Proc. Int. Soc. Magn. Reson. Med.}.
  2009;\hspace{0pt} p. 760.

\bibitem{kim2018fourier}
Kim TH, Haldar JP. The {F}ourier radial error spectrum plot: A more nuanced
  quantitative evaluation of image reconstruction quality. In: \emph{Proc. IEEE
  Int. Symp. Biomed. Imag.}. 2018;\hspace{0pt} pp. 61--64.

\bibitem{mani2019sms_conf}
Mani M, Jacob M, McKinnon G, Yang B, Rutt B, Kerr A, Magnotta V. {SMS-MUSSELS}:
  A navigator-free reconstruction for slice-accelerated multi-shot diffusion
  imaging. In: \emph{Proc. Int. Soc. Magn. Reson. Med.}. 2019;\hspace{0pt} p.
  0233.

\bibitem{bilgic2019}
Bilgic B, Liao C, Manhard MK, Tian Q, Chatnuntawech I, Iyer SS, Cauley SF,
  Feiweier T, Giri S, Y~Hu aYH, Polimeni JR, Wald LL, Setsompop K. Robust
  high-quality multi-shot {EPI} with low-rank prior and machine learning. In:
  \emph{Proc. Int. Soc. Magn. Reson. Med.}. 2019;\hspace{0pt} p. 1250.

\end{thebibliography}

\newpage

\section*{FIGURE AND TABLE CAPTIONS}

\noindent \textbf{Figure 1:} ACS data and reconstruction results for in vivo gradient-echo EPI brain data with an axial slice orientation for different parallel imaging acceleration factors.  Note that the first four acceleration factors ($R=1$-$4$) were acquired from one subject during a single scan session while the last two acceleration factors ($R=5,6$) were acquired from a different subject on a different day, which explains the visual discontinuity between these cases.

\noindent \textbf{Figure 2:} ACS data and reconstruction results for in vivo gradient-echo EPI brain data with a double-oblique slice orientation for different parallel imaging acceleration factors. For reference, we also show the interpolarity phase difference as estimated from a coil-combined RAC-LORAKS result.  The degree of phase nonlinearity is an indicator of how difficult ghost correction is expected to be. As can be seen, complicated 2D nonlinear phase differences are present in many of these cases.

\noindent \textbf{Figure 3:} Reconstruction results for the first set of multi-channel simulations (with similar contrast between ACS and EPI data, but with a hyperintensity added to the EPI data) with different parallel imaging acceleration factors.

\noindent \textbf{Figure 4:} ESPs for the multi-channel simulation results shown in Fig.~3.  The vertical axis of each ESP uses a consistent range to enable comparisons between different acceleration factors.

\noindent \textbf{Figure 5:} Reconstruction results for single-channel simulated data with different  acceleration factors. 

\noindent \textbf{Figure 6:} ACS data and reconstruction results for three representative slices from in vivo diffusion brain data ($R = 3$). A 10$\times$ intensity amplification is also shown for each slice to better highlight the ghosting characteristics. 

\noindent \textbf{Figure 7:} ACS data and reconstruction results for in vivo diffusion EPI brain data for different parallel imaging acceleration factors. For improved visualization, zoomed-in versions of these results (corresponding to the spatial region marked with a yellow rectangle in the first column and first row) are shown in Supporting Information Fig. S9. It should be noted   that the subject appears to have slightly moved between scans, so that there is not perfect correspondence between anatomical image features across different acceleration factors.

\noindent \textbf{Figure 8:} ACS data and reconstruction results for unaccelerated in vivo cardiac EPI data. The two rows show the same results, but the second row has 5$\times$ intensity amplification to better highlight the ghosting characteristics. 

\noindent \textbf{Table 1:} NRMSEs for the multi-channel simulation results shown in Fig.~3. For each acceleration factor, the smallest values are highlighted in bold.

\noindent \textbf{Supporting Information Figure S1:} Illustration of EPI ghost correction. The top row of this figure shows EPI images obtained from different methods, while the bottom row shows the same images with 10$\times$ intensity amplification to highlight the ghost characteristics.  If EPI data is naively reconstructed without accounting for the systematic differences between data acquired with positive and negative readout gradient polarities (``Uncorrected''), then strong Nyquist ghosts appear in the image as indicated with arrows.  Modern EPI techniques frequently try to eliminate these artifacts using navigator information to estimate the systematic differences between the data collected with different readout polarities.  In the navigator-based example we show (``Navigator''), the navigator information was collected using a 3-line EPI acquisition with the phase encoding gradients turned off, and the difference between positive and negative gradient polarities was modeled using constant and 1D linear phase terms.  Although this approach substantially reduces Nyquist ghosts, it is common for some amount of residual ghosting to still be present in the images, particularly in cases where simple 1D phase modeling is inadequate to capture the differences between the two gradient polarities.  We also show an example of our proposed approach (``RAC-LORAKS''), which can account for more complicated variations between the different gradient polarities, and which is substantially more successful at suppressing Nyquist ghosts in this example. 

\noindent \textbf{Supporting Information Figure S2:} Illustration of the orientation of the double-oblique gradient-echo EPI dataset. The double-oblique slices are shown in red, overlaid on a structural T1-weighted image of the same subject. The double-oblique slice used for the results in Fig.~2 is shown with a yellow rectangle.

\noindent \textbf{Supporting Information Figure S3:} Illustration of the EPI and ACS datasets used in simulation.  The first and second top rows show coil-combined multi-channel data for the case when the EPI and ACS data have similar and inverted contrast, respectively, while the bottom row shows  representative single-channel images.   We also show the interpolarity phase difference  for the coil-combined EPI data, as well as the difference in the interpolarity phase difference between the coil-combined EPI and ACS data.  

\noindent \textbf{Supporting Information Figure S4:} The same results shown in Fig.~1, but with a $5\times$ intensity amplification to highlight the ghost characteristics.

\noindent \textbf{Supporting Information Figure S5:} DPG results corresponding to the same data shown in Fig.~1 and Supporting Information Fig.~S4.  The same mDPG results shown in Fig.~1 and Supporting Information Fig.~S4 are also reproduced in this figure for reference. Note that the processing steps of DPG cause the image intensities to be mismatched from the intensities of mDPG and the other reconstruction methods, which precludes a quantitative comparison.  

\noindent \textbf{Supporting Information Figure S6:} mDPG and DPG results corresponding to the same multi-channel simulated data from Fig. 3.

\noindent \textbf{Supporting Information Figure S7:} Reconstruction results for multi-channel simulated data with different parallel imaging acceleration factors. These simulations are identical to those reported in Fig. 3, except that the images used to generate EPI data and the images used to generate ACS data were interchanged. 

\noindent \textbf{Supporting Information Figure S8:}  Reconstruction results for  the second set of multi-channel simulations (with inverted contrast between ACS and EPI data) with different parallel imaging acceleration factors.

\noindent \textbf{Supporting Information Figure S9:} The same results shown in Fig. 7, but zoomed-in to a region of interest for improved visualization.

\noindent \textbf{Supporting Information Table S1:} NRMSEs for the multi-channel inverted contrast simulation results shown in Supporting Information Fig.~S8. For each acceleration factor, the smallest values are highlighted in bold.

\noindent \textbf{Supporting Information Table S2:} NRMSEs for the single-channel simulation results shown in Fig.~5. For each acceleration factor, the smallest values are highlighted in bold.

\clearpage

\begin{table*}[ht]
\caption{NRMSEs for the multi-channel simulation results shown in Fig.~3. For each acceleration factor, the smallest values are highlighted in bold.}
\centering
{\setlength{\tabcolsep}{10pt}
\begin{tabular}{ccccc}
 & \bfseries  MUSSELS & \bfseries mDPG & \bfseries  AC-LORAKS & \bfseries RAC-LORAKS
\\ 
\hline
$R=1$&0.059&0.024&\textbf{0.016}&0.020\\
\rowcolor{lightgray}
$R=2$&0.104&0.045&\textbf{0.035}&0.042\\
$R=3$&0.271&0.083&0.056&\textbf{0.055}\\
\rowcolor{lightgray}
$R=4$&0.572&0.127&0.132&\textbf{0.064}\\
$R=5$&0.741&0.161&0.269&\textbf{0.085}\\
\end{tabular}
}
\label{tab:ret}
\end{table*}

\clearpage

\begin{figure}[ht]
\centering
\includegraphics{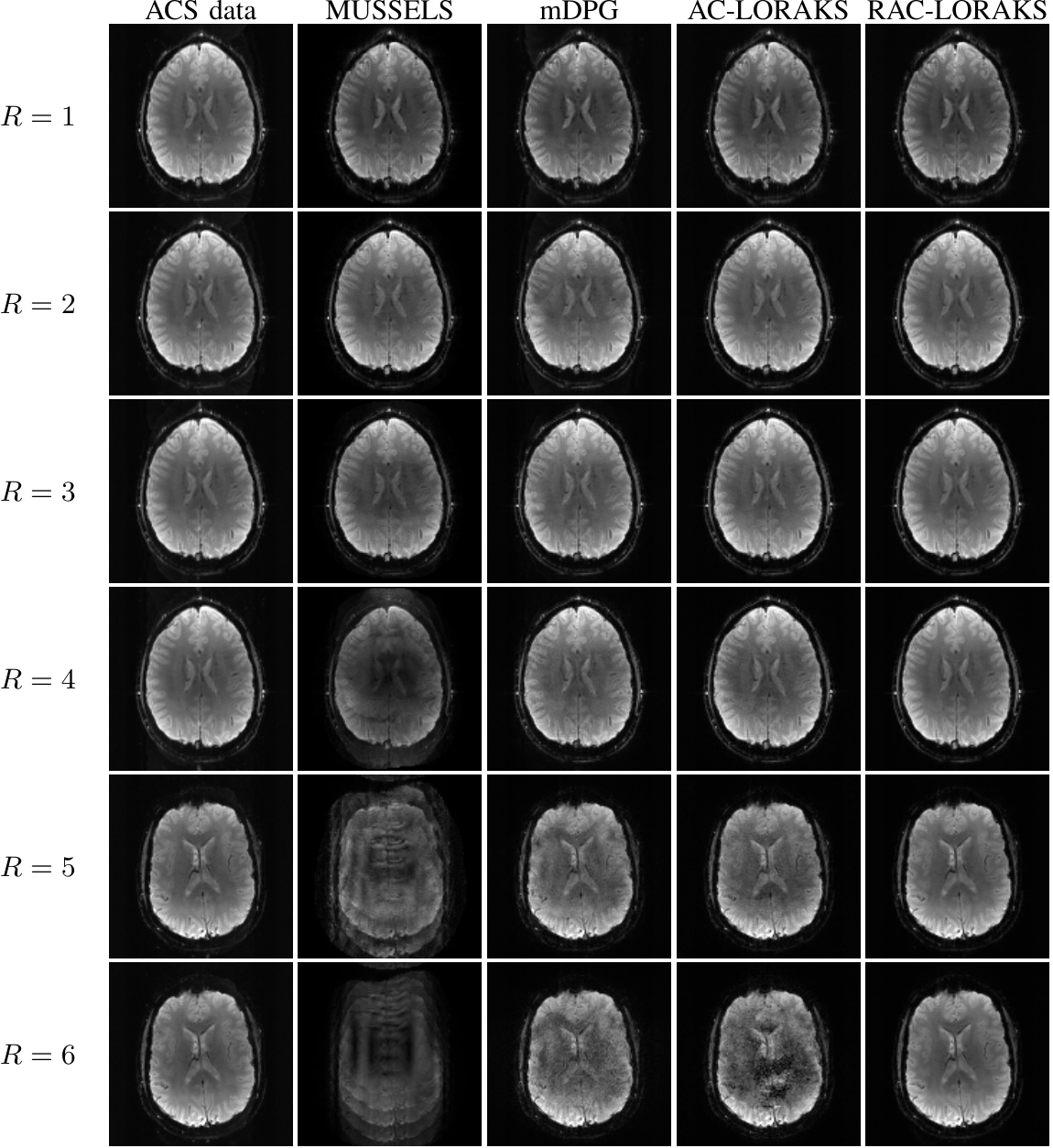}
\caption{ACS data and reconstruction results for in vivo gradient-echo EPI brain data with an axial slice orientation for different parallel imaging acceleration factors.  Note that the first four acceleration factors ($R=1$-$4$) were acquired from one subject during a single scan session while the last two acceleration factors ($R=5,6$) were acquired from a different subject on a different day, which explains the visual discontinuity between these cases.} 
\label{fig:pros_axial}			
\end{figure}

\clearpage

\begin{figure}[ht]
\centering
\includegraphics{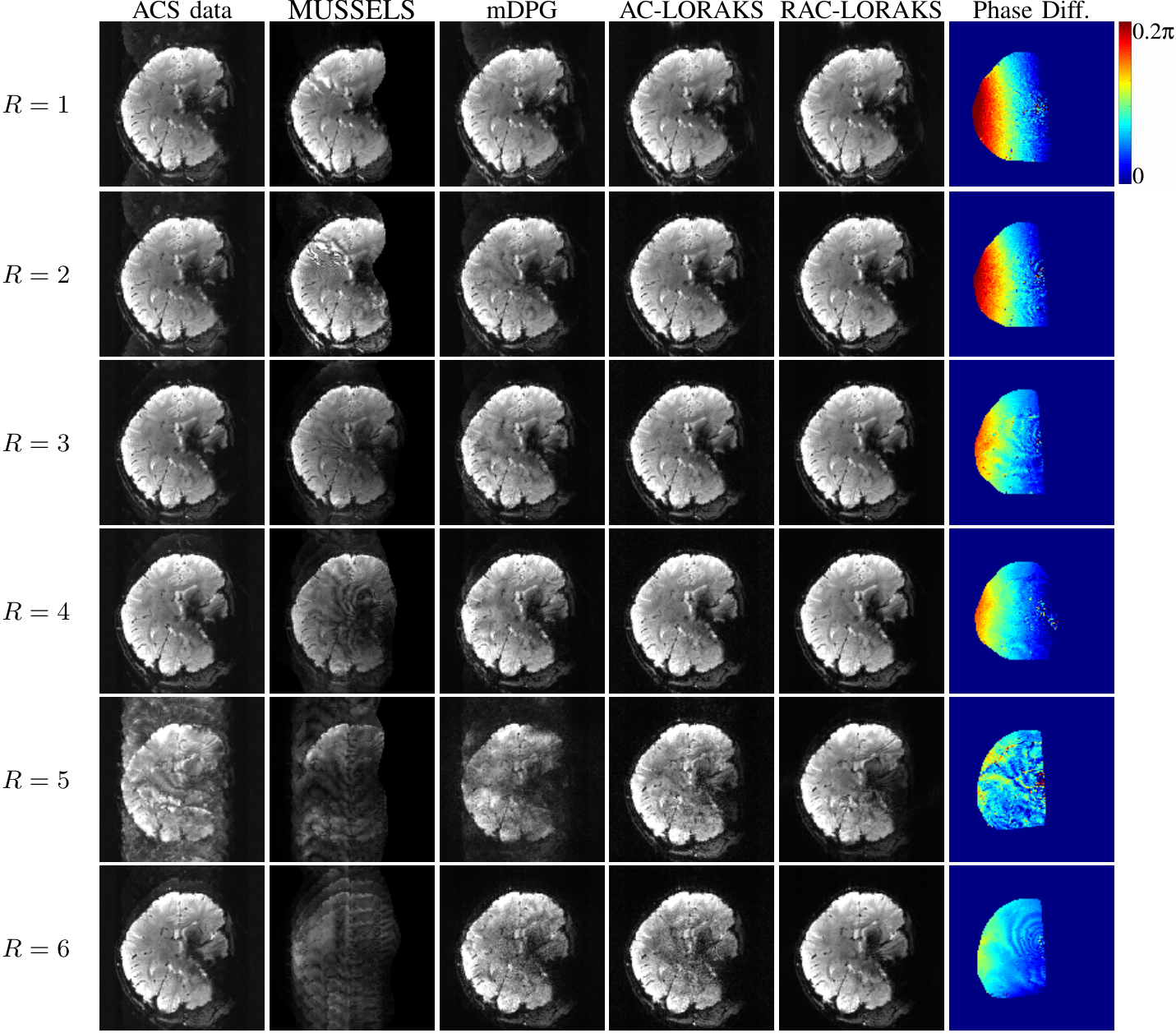}
\caption{ACS data and reconstruction results for in vivo gradient-echo EPI brain data with a double-oblique slice orientation for different parallel imaging acceleration factors. For reference, we also show the interpolarity phase difference as estimated from a coil-combined RAC-LORAKS result.  The degree of phase nonlinearity is an indicator of how difficult ghost correction is expected to be. As can be seen, complicated 2D nonlinear phase differences are present in many of these cases.
} 
\label{fig:pros_doblique}		
\end{figure}

\clearpage

\begin{figure}[ht]
\centering
\includegraphics{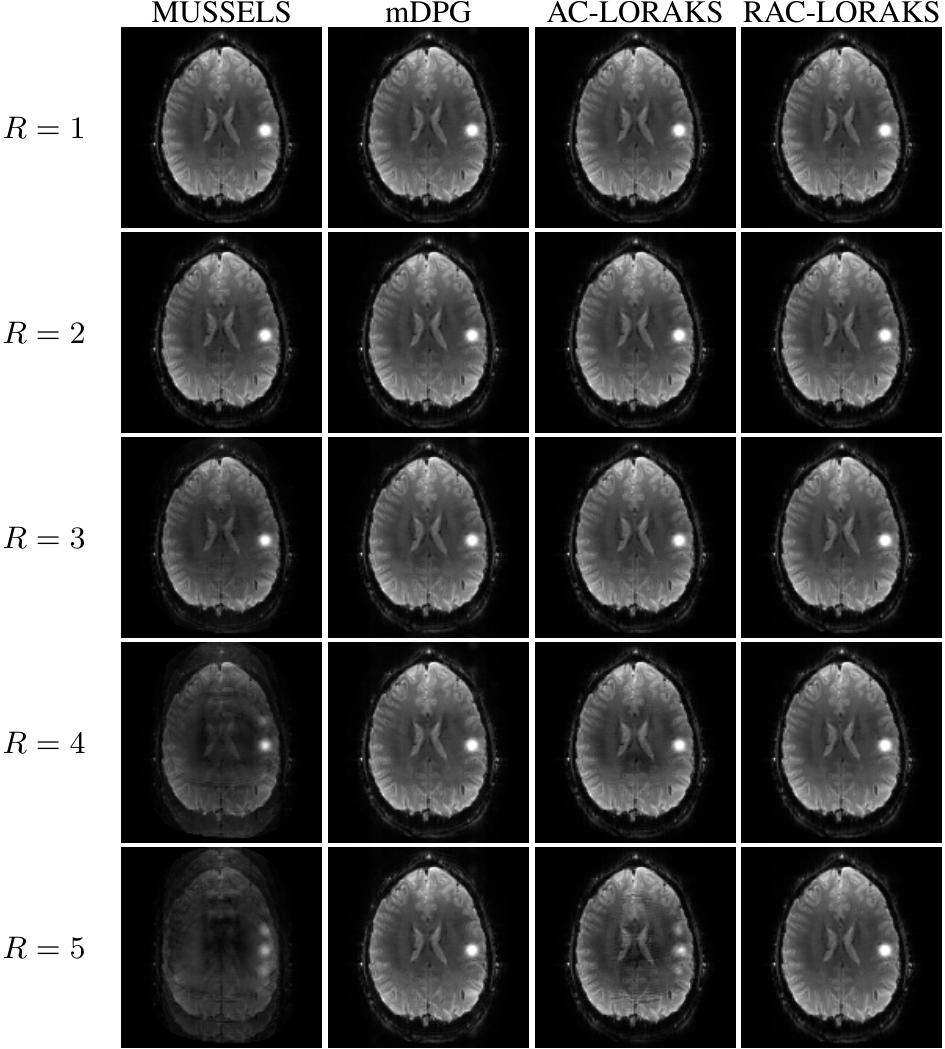}
\caption{Reconstruction results for  the first set of multi-channel simulations (with similar contrast between ACS and EPI data, but with a hyperintensity added to the EPI data) with different parallel imaging acceleration factors. } 
\label{fig:ret}			
\end{figure}


\clearpage

\begin{figure}[ht]
\centering
\includegraphics[width=0.6\textwidth]{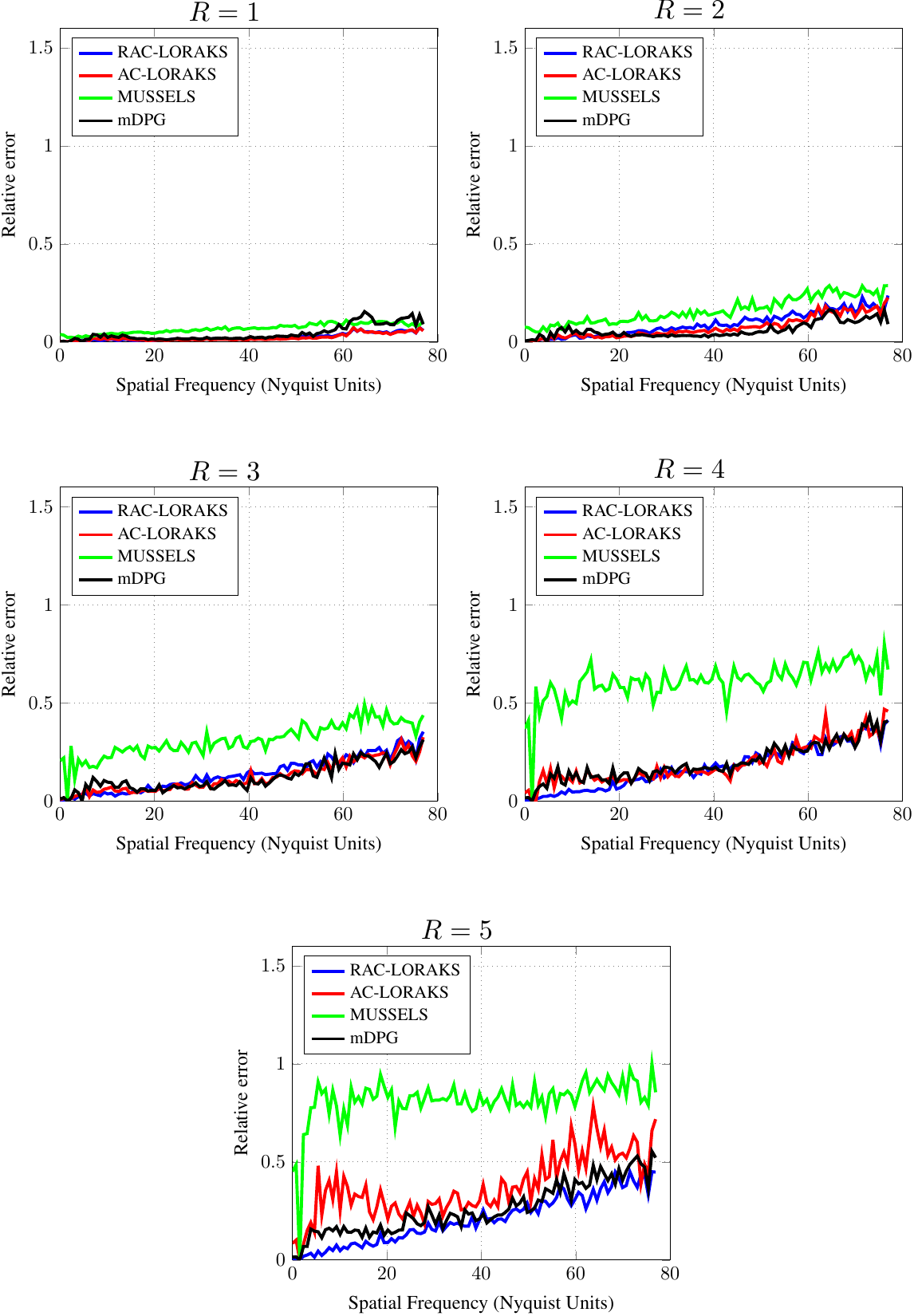}
\caption{ESPs for the multi-channel simulation results shown in Fig.~3.  The vertical axis of each ESP uses a consistent range to enable comparisons between different acceleration factors.
}   
\label{fig:esp_plots}
\end{figure}

\clearpage

\begin{figure}[ht]
\centering
\includegraphics{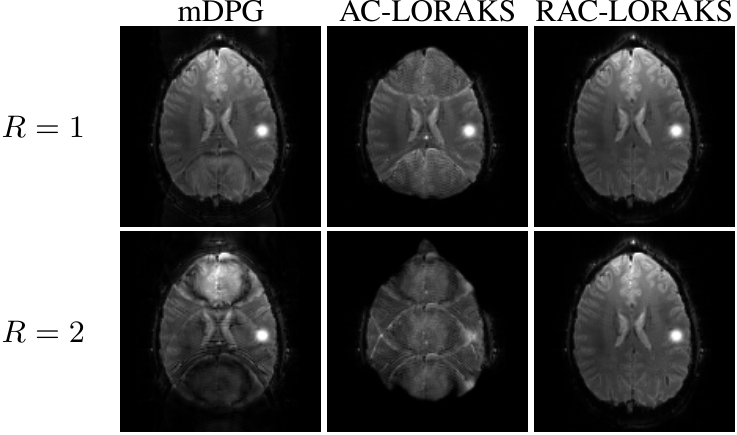}
\caption{Reconstruction results for single-channel simulated data with different  acceleration factors. } 
\label{fig:ret_sc}		
\end{figure}

\clearpage
\begin{figure}[ht]
\centering
\includegraphics{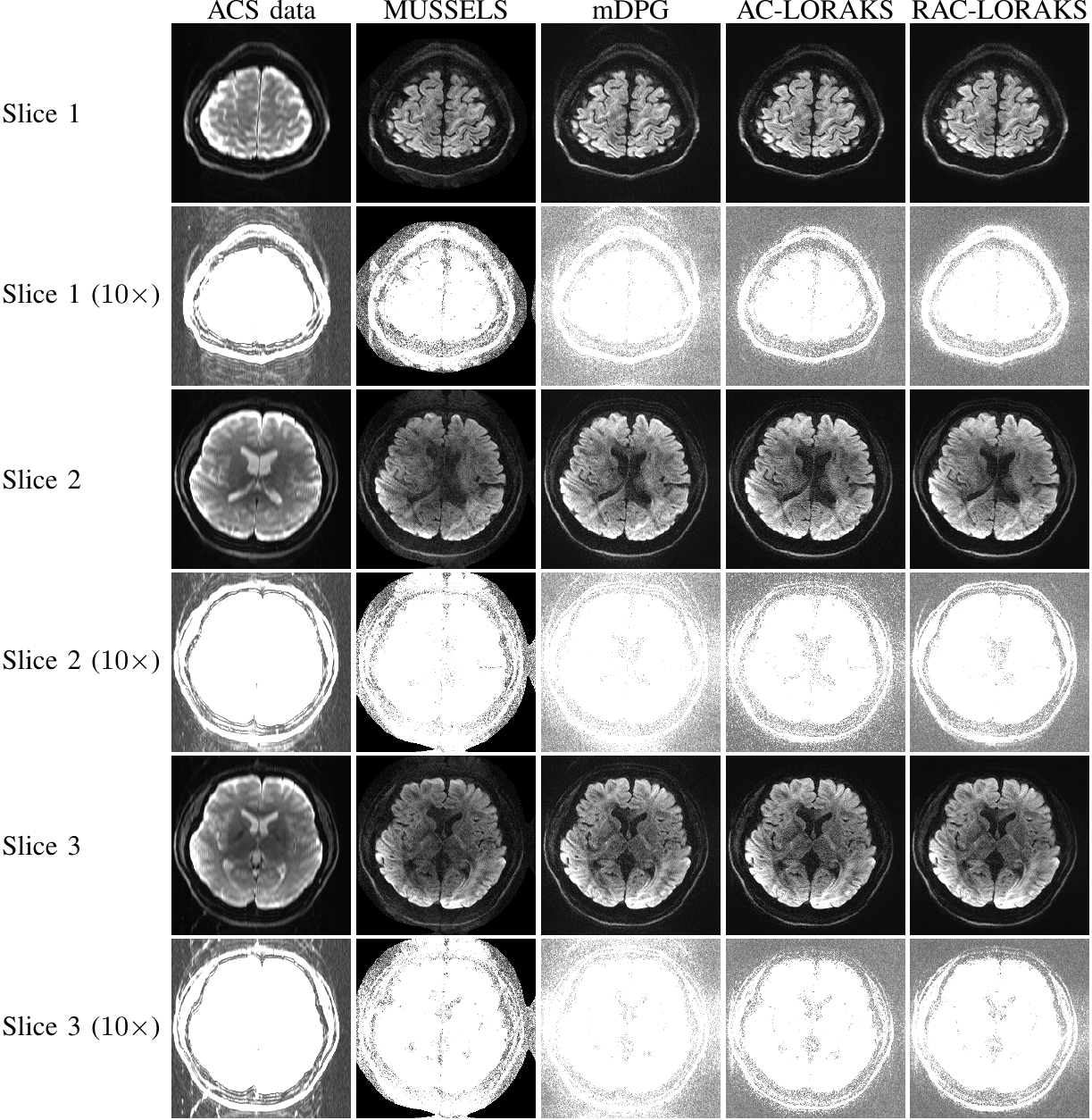}
\caption{ACS data and reconstruction results for three representative slices from in vivo diffusion brain data ($R = 3$). A 10$\times$ intensity amplification is also shown for each slice to better highlight the ghosting characteristics. } 
\label{fig:pros_diff}		
\end{figure}

\clearpage

\begin{figure}[ht]
\centering
\includegraphics[width=0.8\textwidth]{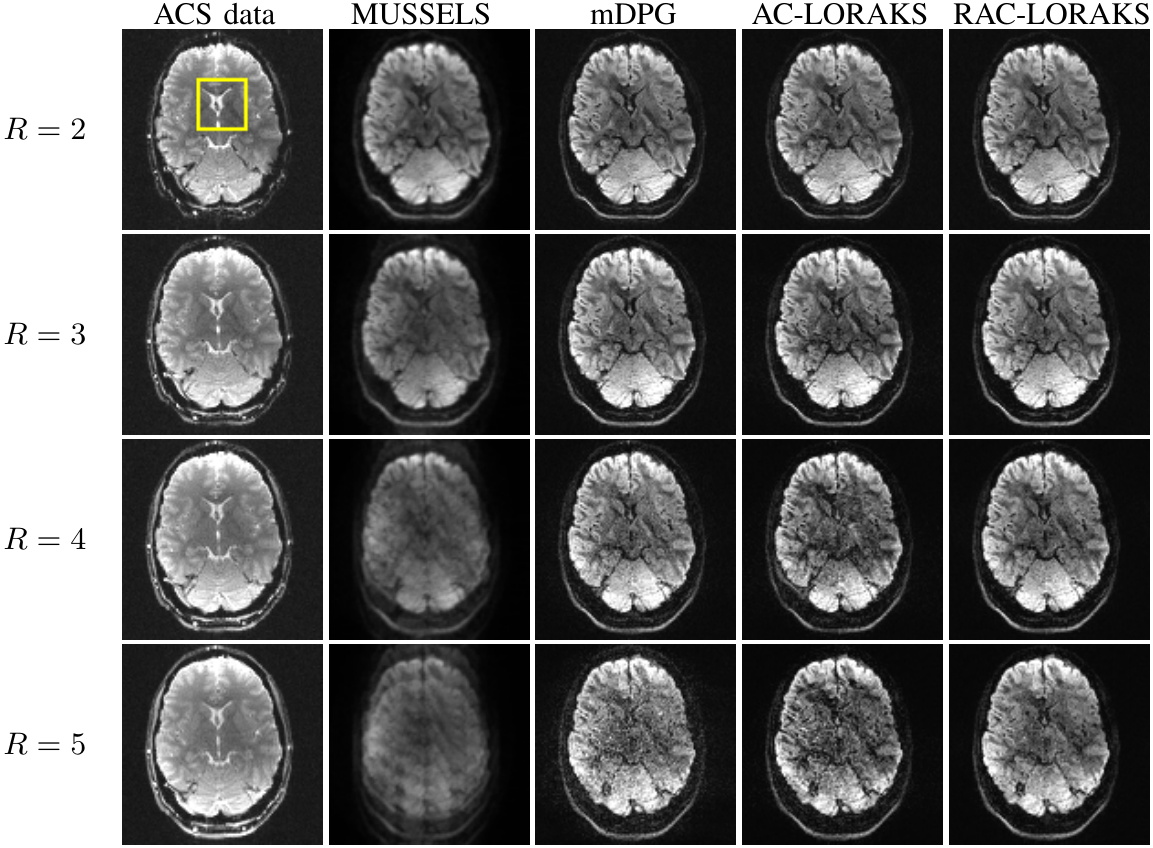}
\caption{ACS data and reconstruction results for in vivo diffusion EPI brain data for different parallel imaging acceleration factors. For improved visualization, zoomed-in versions of these results (corresponding to the spatial region marked with a yellow rectangle in the first column and first row) are shown in Supporting Information Fig. S9. It should be noted   that the subject appears to have slightly moved between scans, so that there is not perfect correspondence between anatomical image features across different acceleration factors.} 
\label{fig:pros_diff_sup}	
\end{figure}

\clearpage
\begin{figure}[ht]
\centering
\includegraphics{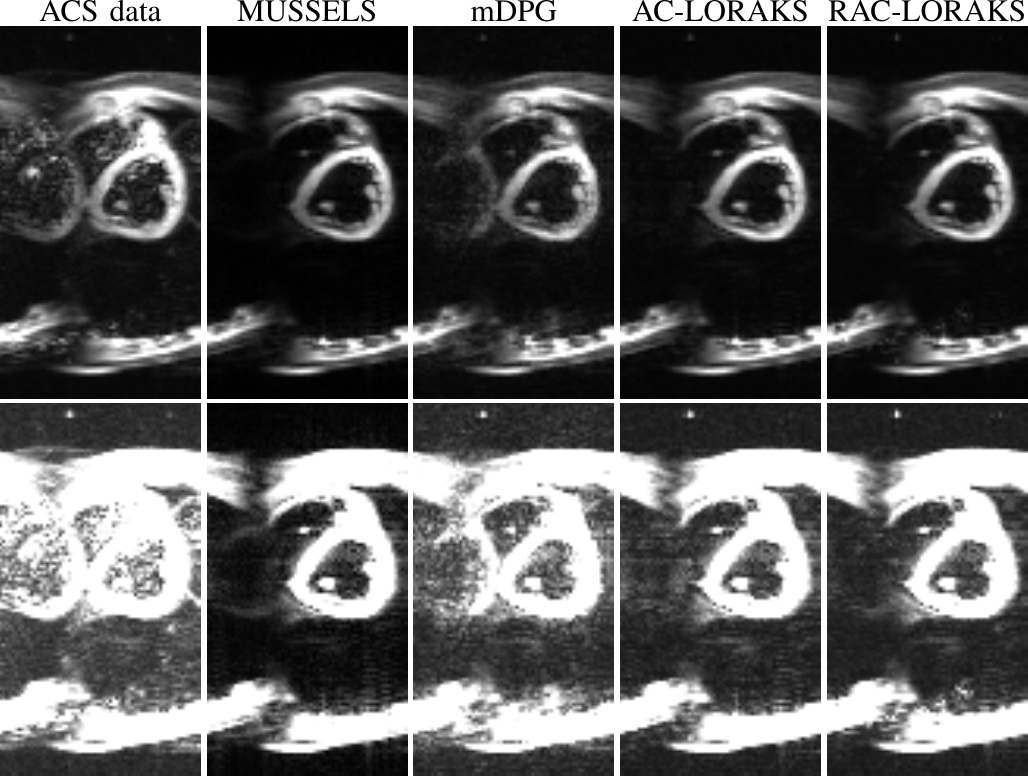}
\caption{ACS data and reconstruction results for unaccelerated in vivo cardiac EPI data. The two rows show the same results, but the second row has 5$\times$ intensity amplification to better highlight the ghosting characteristics. } 
\label{fig:pros_cardiac}		
\end{figure}

\clearpage 

\setcounter{figure}{0}
\setcounter{table}{0}
\makeatletter 
\renewcommand{\thefigure}{S\arabic{figure}}
\renewcommand{\thetable}{S\arabic{table}}

\section*{SUPPORTING INFORMATION } 

\begin{figure}[ht]
\centering
\includegraphics{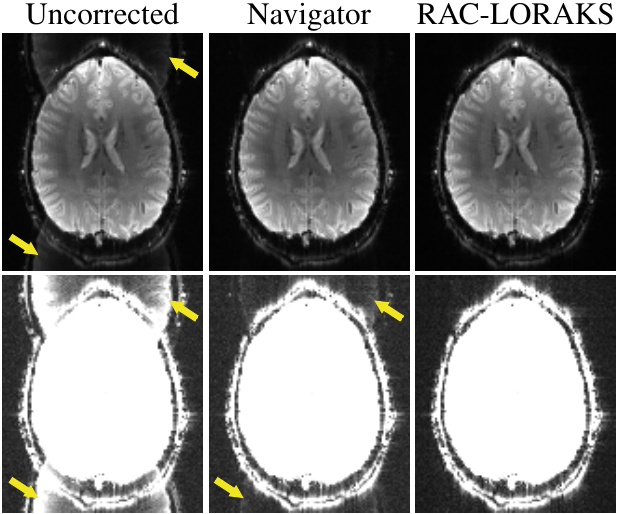}
\caption*{Supporting Information Figure S1: Illustration of EPI ghost correction. The top row of this figure shows EPI images obtained from different methods, while the bottom row shows the same images with 10$\times$ intensity amplification to highlight the ghost characteristics.  If EPI data is naively reconstructed without accounting for the systematic differences between data acquired with positive and negative readout gradient polarities (``Uncorrected''), then strong Nyquist ghosts appear in the image as indicated with arrows.  Modern EPI techniques frequently try to eliminate these artifacts using navigator information to estimate the systematic differences between the data collected with different readout polarities.  In the navigator-based example we show (``Navigator''), the navigator information was collected using a 3-line EPI acquisition with the phase encoding gradients turned off, and the difference between positive and negative gradient polarities was modeled using constant and 1D linear phase terms.  Although this approach substantially reduces Nyquist ghosts, it is common for some amount of residual ghosting to still be present in the images, particularly in cases where simple 1D phase modeling is inadequate to capture the differences between the two gradient polarities.  We also show an example of our proposed approach (``RAC-LORAKS''), which can account for more complicated variations between the different gradient polarities, and which is substantially more successful at suppressing Nyquist ghosts in this example.  }

\label{fig:nav}			
\end{figure}

\clearpage

\begin{figure}[ht]
\centering
\includegraphics[width=0.8\textwidth]{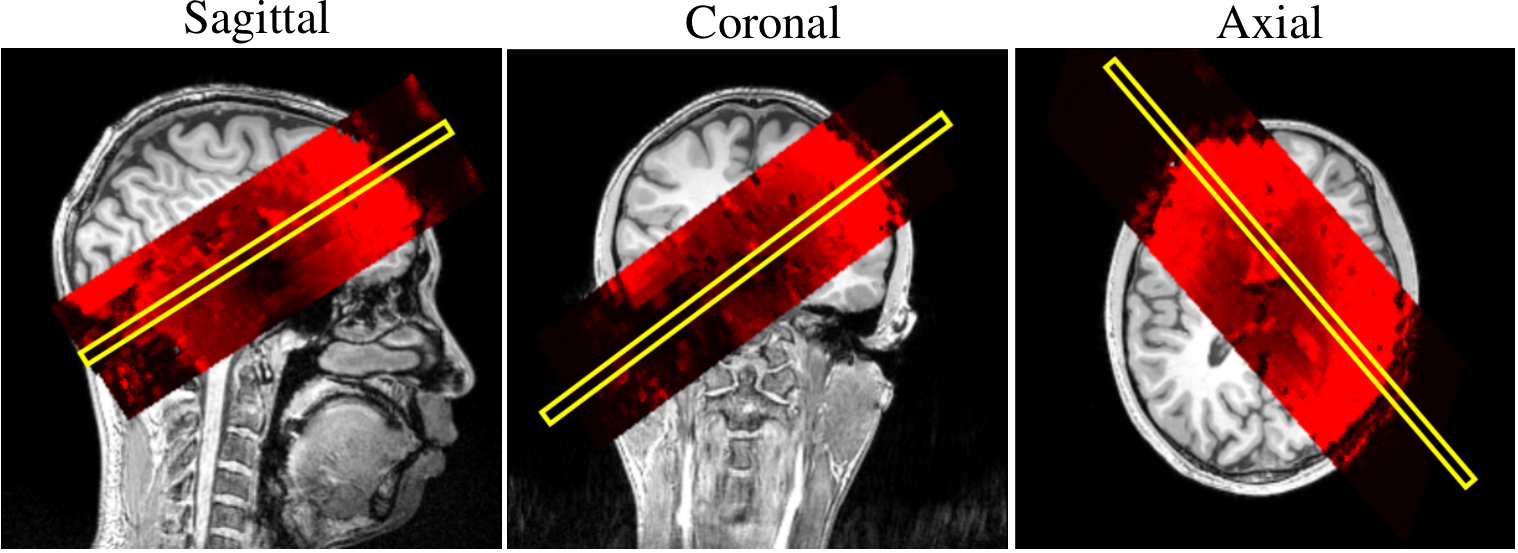}
\caption*{Supporting Information Figure S2: Illustration of the orientation of the double-oblique gradient-echo EPI dataset. The double-oblique slices are shown in red, overlaid on a structural T1-weighted image of the same subject. The double-oblique slice used for the results in Fig.~2 is shown with a yellow rectangle.
}
\label{fig:sim}			
\end{figure}

\clearpage

\begin{figure}[ht]
\centering
\includegraphics{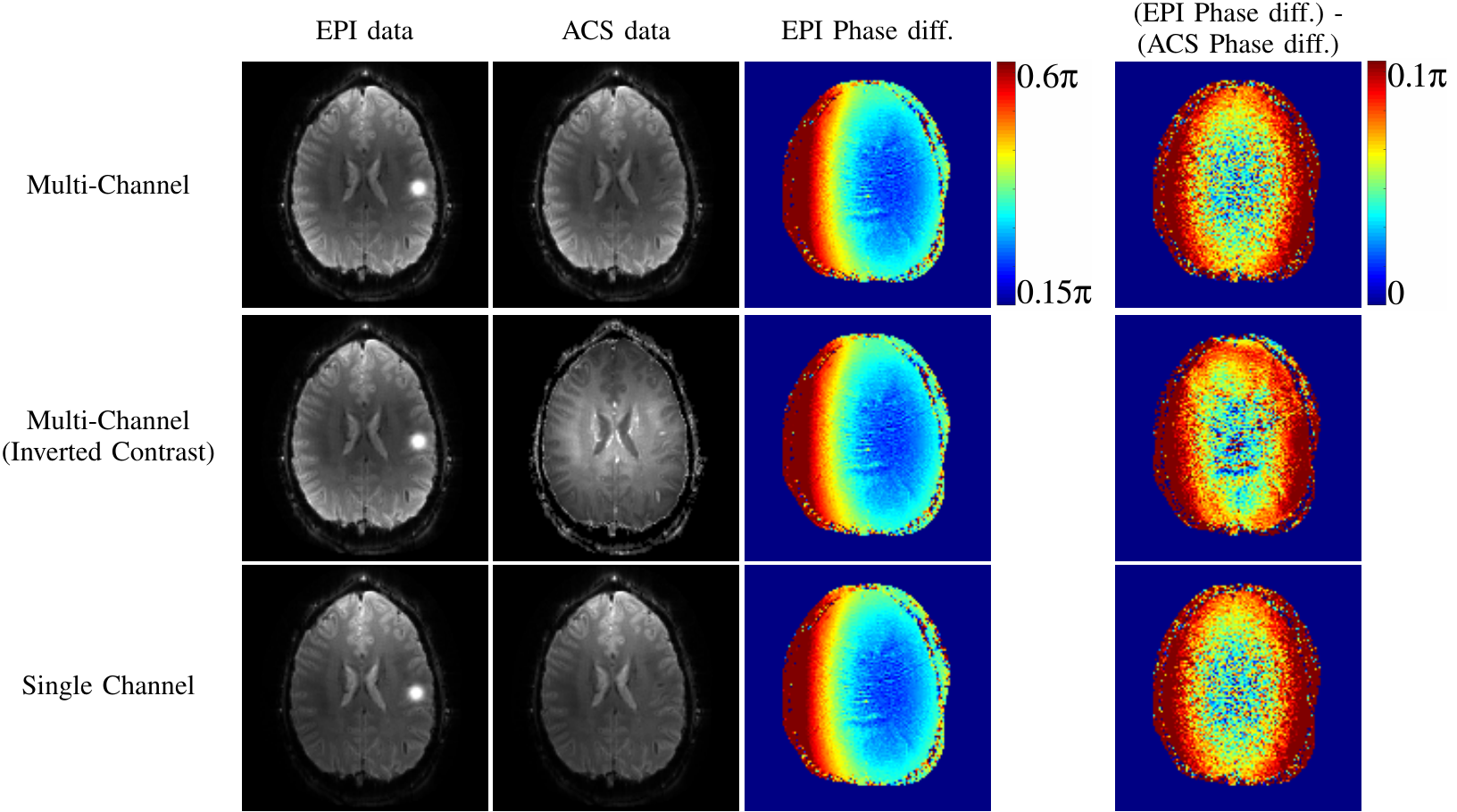}
\caption*{Supporting Information Figure S3: Illustration of the EPI and ACS datasets used in simulation.  The first and second top rows show coil-combined multi-channel data for the case when the EPI and ACS data have similar and inverted contrast, respectively, while the bottom row shows  representative single-channel images.   We also show the interpolarity phase difference  for the coil-combined EPI data, as well as the difference in the interpolarity phase difference between the coil-combined EPI and ACS data. 
}
\label{fig:sim}			
\end{figure}

\clearpage 

\begin{figure}[ht]
\centering
\includegraphics{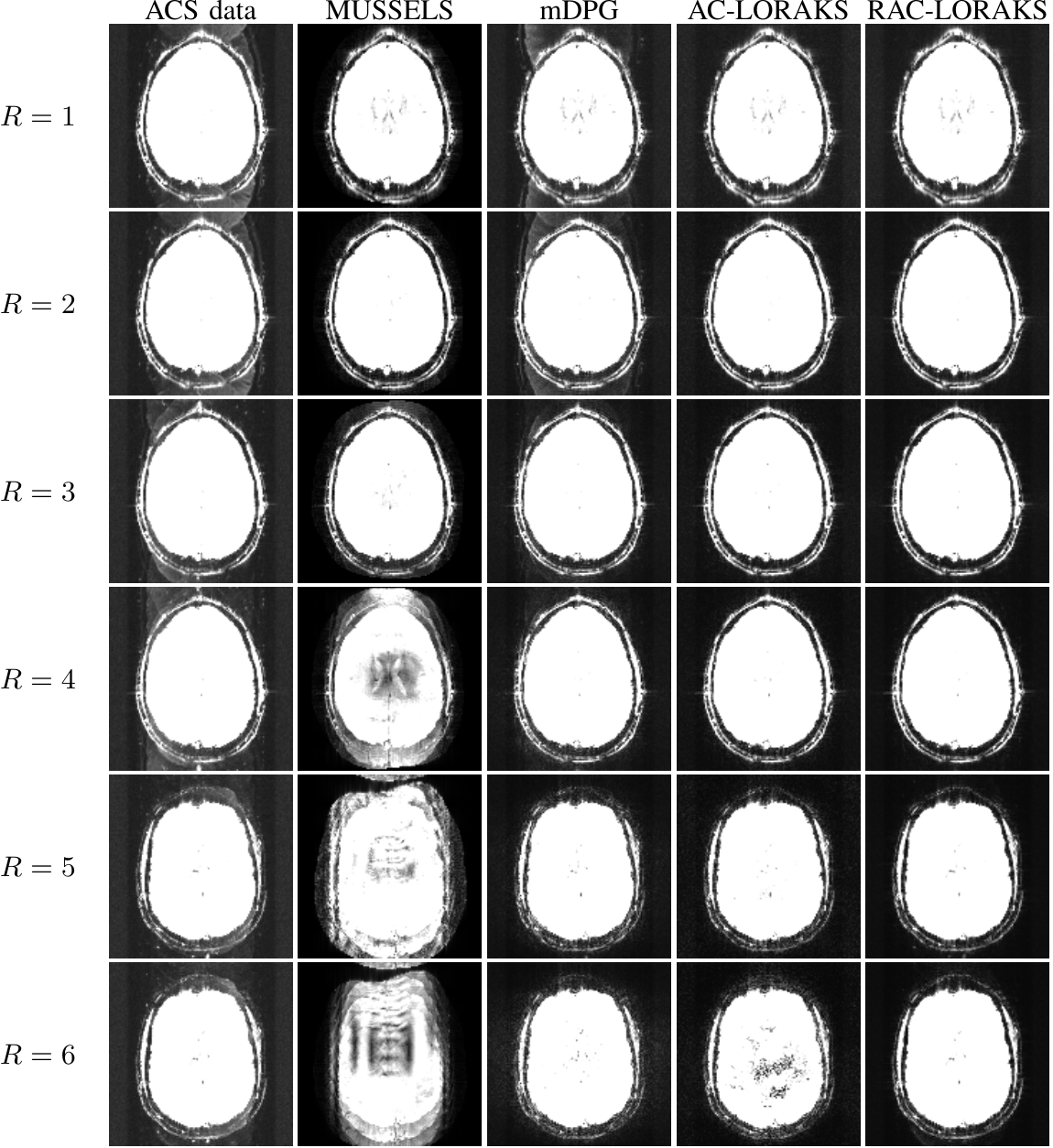}
\caption*{Supporting Information Figure S4: The same results shown in Fig.~1, but with a $5\times$ intensity amplification to highlight the ghost characteristics.} 
\label{fig:axial_W2}			
\end{figure}

\clearpage 

\begin{figure}[ht]
\centering
\includegraphics{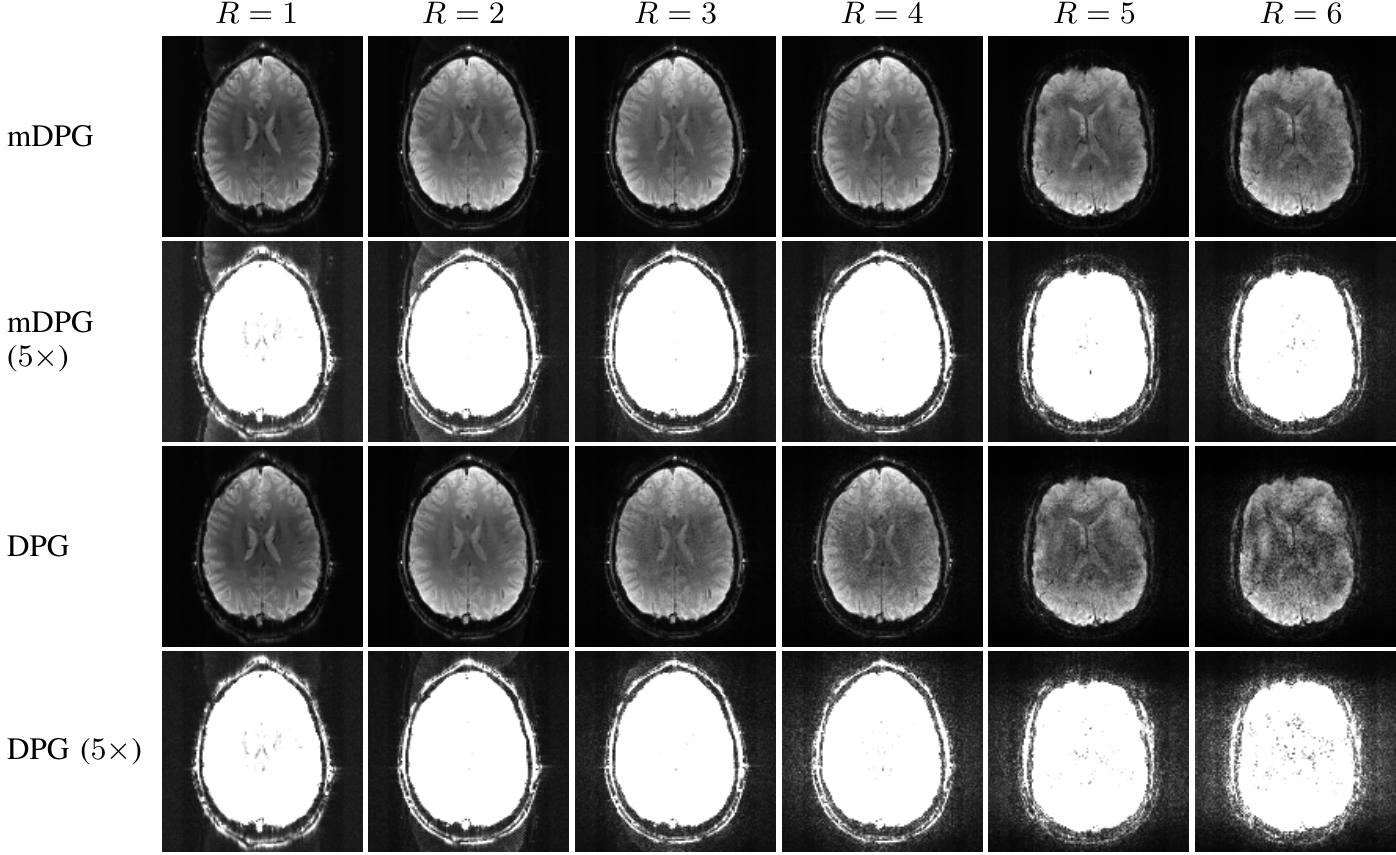}
\caption*{ Supporting Information Figure S5: DPG results corresponding to the same data shown in Fig.~1 and Supporting Information Fig.~S4.  The same mDPG results shown in Fig.~1 and Supporting Information Fig.~S4 are also reproduced in this figure for reference. Note that the processing steps of DPG cause the image intensities to be mismatched from the intensities of mDPG and the other reconstruction methods, which precludes a quantitative comparison.  
}
\label{fig:axial_DPG_ori}			
\end{figure}

\clearpage 

\begin{figure}[ht]
\centering
\includegraphics{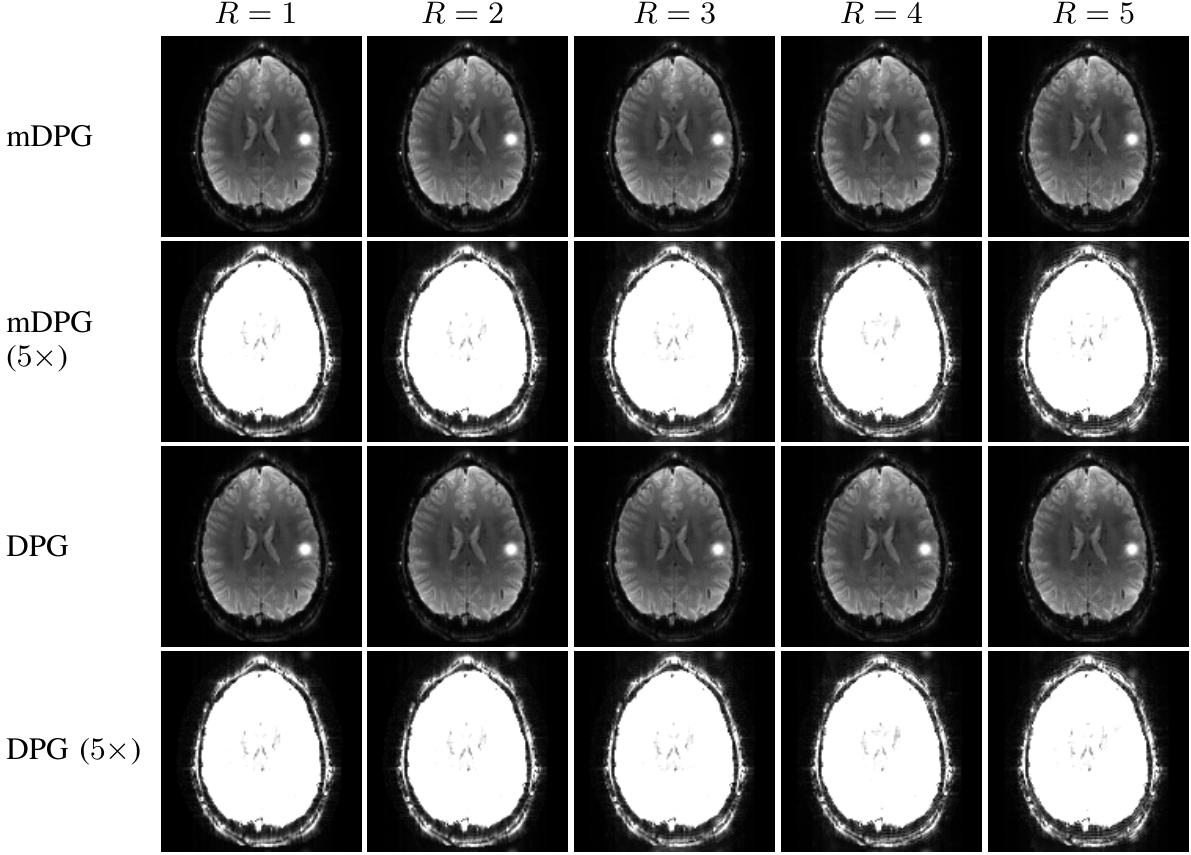}
\caption*{Supporting Information Figure S6: mDPG and DPG results corresponding to the same multi-channel simulated data from Fig. 3.}
\label{fig:axial_DPG_ori}			
\end{figure}

\clearpage

\begin{figure}[ht]
\centering
\includegraphics{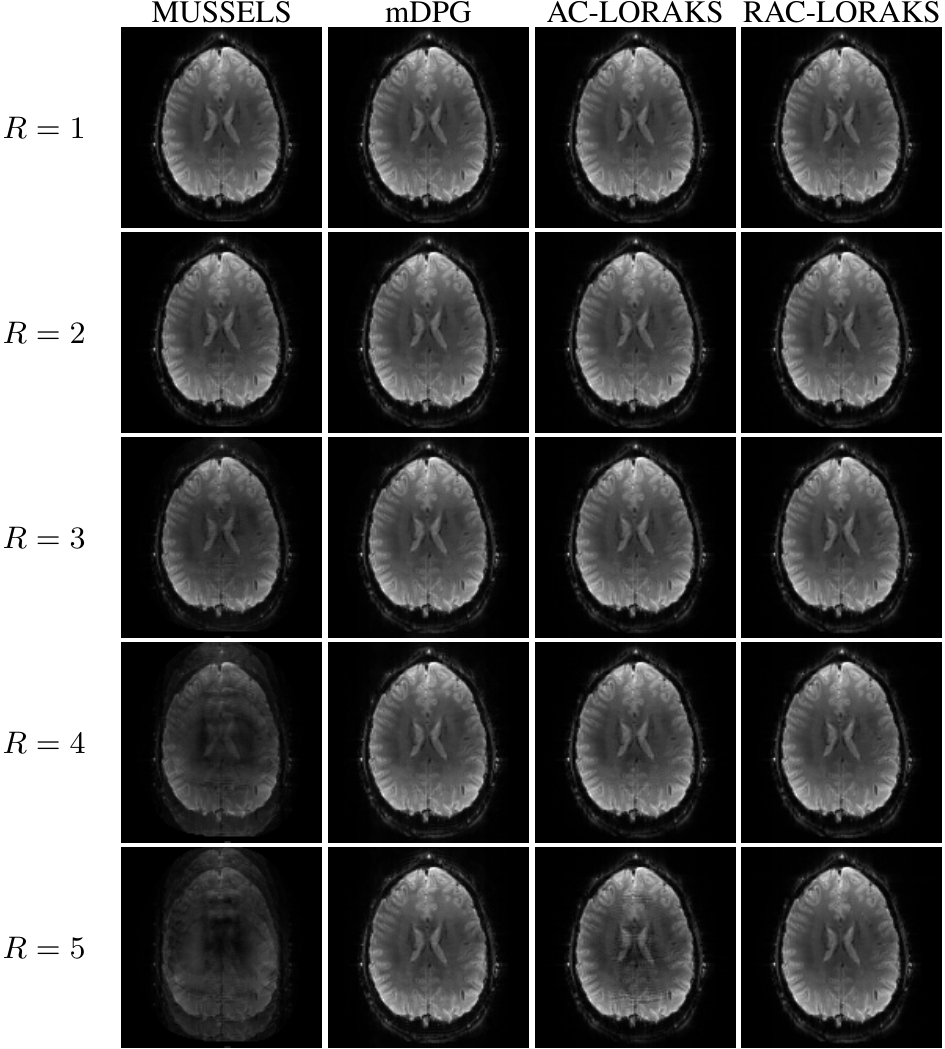}
\caption*{ Supporting Information Figure S7: Reconstruction results for multi-channel simulated data with different parallel imaging acceleration factors. These simulations are identical to those reported in Fig. 3, except that the images used to generate EPI data and the images used to generate ACS data were interchanged. 
}
\label{fig:inverted}			
\end{figure}

\clearpage

\begin{figure}[ht]
\centering
\includegraphics{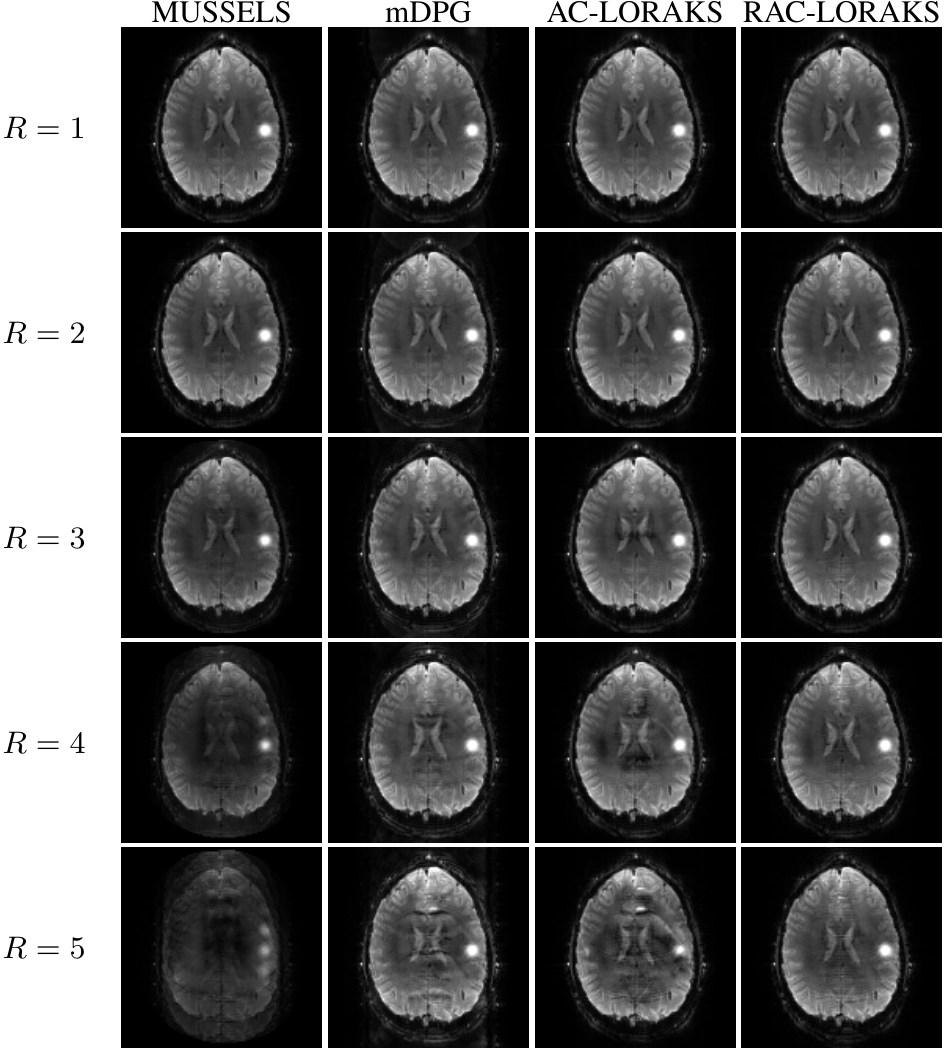}
\caption*{ Supporting Information Figure S8:  Reconstruction results for  the second set of multi-channel simulations (with inverted contrast between ACS and EPI data) with different parallel imaging acceleration factors.
}
\label{fig:dif_contrast}			
\end{figure}

\clearpage 

\begin{figure}[ht]
\centering
\includegraphics{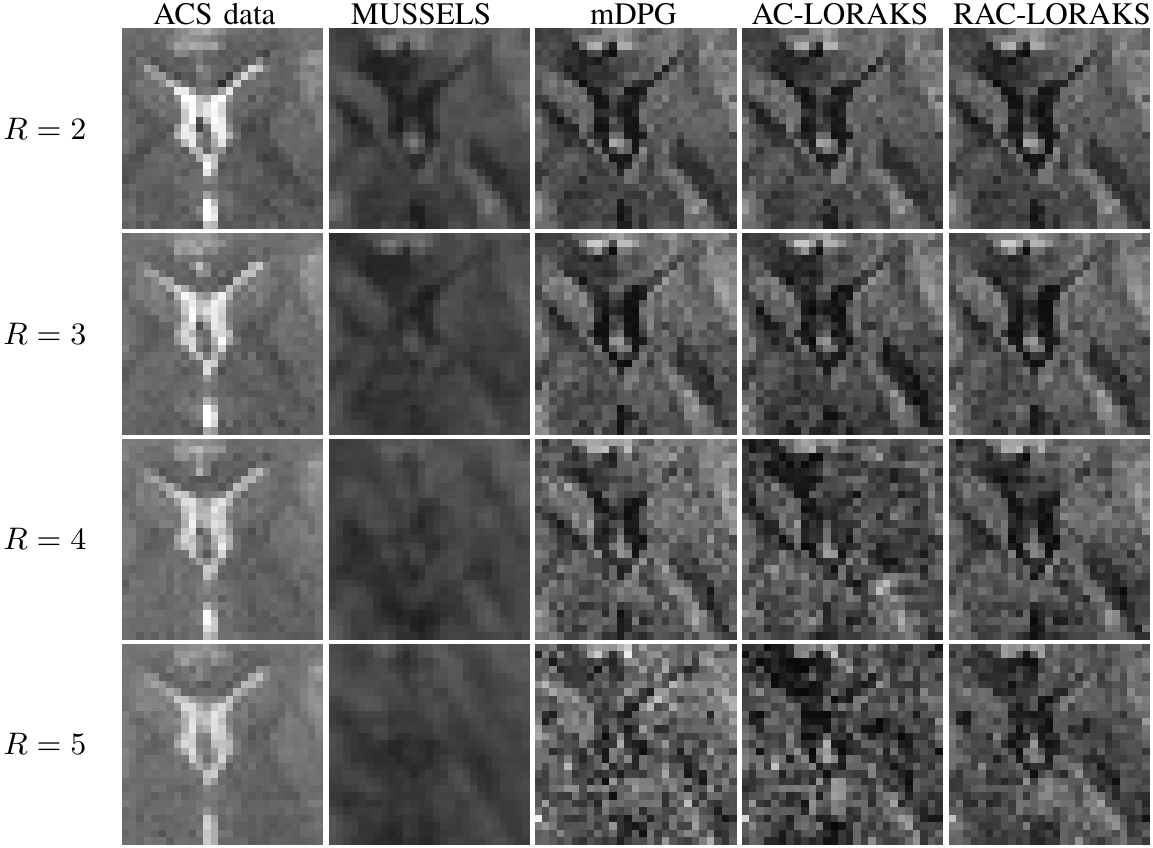}
\caption*{ Supporting Information Figure S9: The same results shown in Fig. 7, but zoomed-in to a region of interest for improved visualization.} 
\label{fig:inverted}			
\end{figure}

\clearpage

\begin{figure}[ht]
\centering
\includegraphics{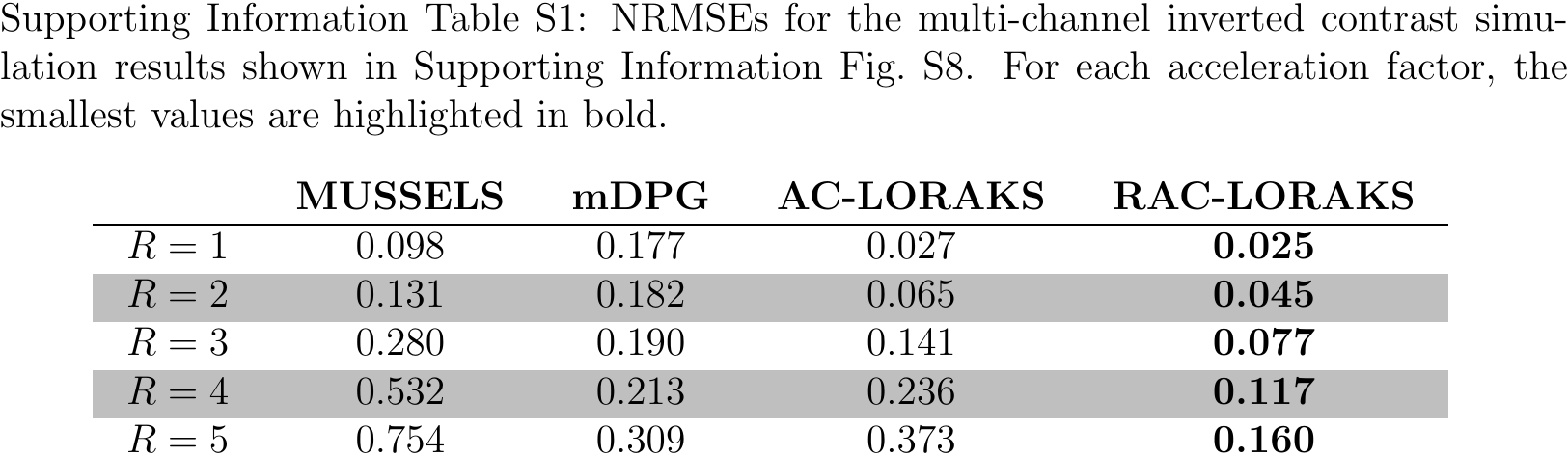}
\end{figure}

\clearpage

\begin{figure}[ht]
\centering
\includegraphics{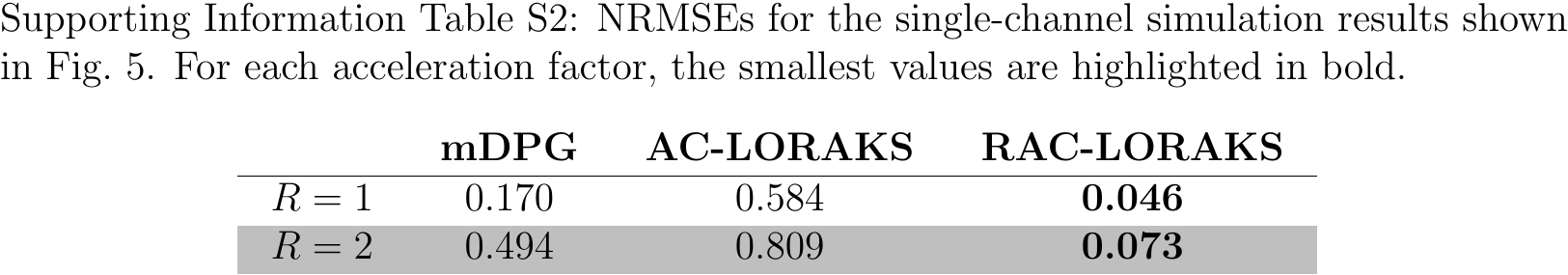}
\end{figure}

\end{document}